\documentclass[12pt,preprint]{aastex}
\usepackage{epsfig}
\usepackage{rotating}
\usepackage{comment}

\def\etal{{et al.\thinspace}}

\def\spose#1{\hbox to 0pt{#1\hss}}

\def\multleft#1{\hbox to size{\vbox {\halign {\lft{##}\cr #1}}\hfill}\par}
\def\multright#1{\hbox to size{\vbox {\halign {\rt{##}\cr #1}}\hfill}\par}

\def\degmark{^\circ}
\def\boxit#1{\vbox{\hrule\hbox{\vrule\kern3pt\vbox{\kern3pt
          #1 \kern3pt}\kern3pt\vrule}\hrule}}

\def\cm{{\rm\thinspace cm}}

\def\erg{{\rm\thinspace erg}}
\def\eV{{\rm\thinspace eV}}

\def\keV{{\rm\thinspace keV}}
\def\km{{\rm\thinspace km}}

\def\Msun{\hbox{$\rm\thinspace M_{\odot}$}}

\def\pc{{\rm\thinspace pc}}
\def\ph{{\rm\thinspace ph}}
\def\s{{\rm\thinspace s}}
\def\ks{{\rm\thinspace ks}}

\def\cts{{\rm\thinspace cts}}

\def\ergcmps{\hbox{$\erg\cm\s^{-1}\,$}}

\def\ergpcmsqps{\hbox{$\erg\cm^{-2}\s^{-1}\,$}}

\def\ergps{\hbox{$\erg\s^{-1}\,$}}

\def\kmps{\hbox{$\km\s^{-1}\,$}}

\def\pcmsq{\hbox{$\cm^{-2}\,$}}

\def\phpcmsqps{\hbox{$\ph\cm^{-2}\s^{-1}\,$}}

\def\ctsps{\hbox{$\cts\s^{-1}$}}

\makeatletter

\let\@internalcite\cite
\def\cite{\@ifstar{\citey}{\citefull}}
\def\citefull{\def\astroncite##1##2{##1\ ##2}\@internalcite}
\def\citey{\def\astroncite##1##2{##1\ (##2)}\@internalcite}
\def\citeyear{\def\astroncite##1##2{##2}\@internalcite}
\def\citename{\def\astroncite##1##2{##1}\@internalcite}
\def\@citex[#1]#2{\if@filesw\immediate\write\@auxout{\string\citation{#2}}\fi
  \def\@citea{}\@cite{\@for\@citeb:=#2\do
    {\@citea\def\@citea{; }\@ifundefined
       {b@\@citeb}{{\bf ??}\@warning
       {Citation `\@citeb' on page \thepage \space undefined}}%
{\csname b@\@citeb\endcsname}}}{#1}}

\def\@cite#1#2{#1\if@tempswa #2\fi}
\def\@biblabel#1{}

\def\astroncite#1#2{#1\ #2}
\makeatother

\begin{document}

\title{The Broad-band X-ray Spectrum of
  IC~4329A from a Joint {\em NuSTAR/Suzaku} Observation}

\author{L.~W.~Brenneman\altaffilmark{1}, 
G.~Madejski\altaffilmark{2},
F.~Fuerst\altaffilmark{3},
G.~Matt\altaffilmark{4},
M.~Elvis\altaffilmark{1},
F.~A.~Harrison\altaffilmark{3},
D.~R.~Ballantyne\altaffilmark{5},
S.~E.~Boggs\altaffilmark{6},
F.~E.~Christensen\altaffilmark{7},
W.~W.~Craig\altaffilmark{7,8},
A.~C.~Fabian\altaffilmark{9},
B.~W.~Grefenstette\altaffilmark{3},
C.~J.~Hailey\altaffilmark{10},
K.~K.~Madsen\altaffilmark{3},
A.~Marinucci\altaffilmark{4},
E.~Rivers\altaffilmark{3},
D.~Stern\altaffilmark{11},
D.~J.~Walton\altaffilmark{3},
W.~W.~Zhang\altaffilmark{12}}
\altaffiltext{1}{Harvard-Smithsonian CfA, 60 Garden St. MS-67,
  Cambridge, MA 02138, USA}
\altaffiltext{2}{Kavli Institute for Particle Astrophysics and
  Cosmology, SLAC National Accelerator Laboratory, Menlo Park, CA 94025, USA}
\altaffiltext{3}{Cahill Center for Astronomy and Astrophysics,
  California Institute of Technology, Pasadena, CA 91125, USA}
\altaffiltext{4}{Dipartimento di Matematica e Fisica, Universit\`{a} Roma
  Tre, via della Vasca Navale 84, I-00146 Roma, Italy}
\altaffiltext{5}{Center for Relativistic Astrophysics, School of
  Physics, Georgia Institute of Technology, Atlanta, GA 30332, USA}
\altaffiltext{6}{Space Science Laboratory, University of California, Berkeley,
  California 94720, USA}
\altaffiltext{7}{DTU Space—National Space Institute, Technical University of
  Denmark, Elektrovej 327, 2800 Lyngby, Denmark}
\altaffiltext{8}{Lawrence Livermore National Laboratory, Livermore, California
  94550, USA}
\altaffiltext{9}{Institute of Astronomy, Madingley Road, Cambridge CB3 0HA, UK}
\altaffiltext{10}{Columbia Astrophysics Laboratory, Columbia University, New
  York, New York 10027, USA}
\altaffiltext{11}{Jet Propulsion Laboratory, California Institute of
  Technology, Pasadena, CA 91109, USA} 
\altaffiltext{12}{NASA Goddard Space Flight Center, Greenbelt, Maryland 20771, USA}

\begin{abstract}
\noindent 
We have obtained a deep, simultaneous observation of the bright, nearby Seyfert
galaxy IC~4329A with {\it
  Suzaku} and {\it NuSTAR}.  Through a detailed spectral analysis,
we are able to robustly separate the continuum, absorption and
distant reflection components in the spectrum.  The absorbing column is found to be modest
($\sim6 \times 10^{21} \pcmsq$), and does not introduce any
significant curvature in the Fe K
band.  We are able to place a strong constraint on the
presence of a broadened Fe K$\alpha$ line ($E_{\rm
  rest}=6.46^{+0.08}_{-0.07} \keV$ with $\sigma=0.33^{+0.08}_{-0.07}
\keV$ and $EW=34^{+8}_{-7} \eV$), though we are not able to
constrain any of the parameters of a relativistic reflection model.  
These results highlight the range in broad Fe K line strengths
observed in nearby, bright AGN (roughly an order of magnitude), and imply a corresponding range in
the physical properties of the inner accretion disk in these sources.
We have also updated our previously reported
measurement of the high-energy cutoff of the hard X-ray emission using
both observatories rather than just {\it NuSTAR} alone: $E_{\rm
  cut}=186 \pm 14 \keV$.  This
high-energy cutoff acts as
a proxy for the temperature of the coronal electron plasma, enabling us to further
separate this parameter from the plasma's optical depth and to update
our results for these parameters as well.  We derive $kT=50^{+6}_{-3}
\keV$ with $\tau=2.34^{+0.16}_{-0.11}$ using a spherical geometry,
$kT=61 \pm 1 \keV$ with $\tau=0.68 \pm 0.02$ for a slab
geometry, with both having an equivalent goodness-of-fit.
\end{abstract}

\section{Introduction}
\label{sec:intro}

X-ray observations of active galactic nuclei (AGN) elucidate many
physical processes that drive the production of high-energy photons
close to a supermassive black hole (SMBH).  In addition to probing the
properties of the corona in radio-quiet AGN, X-ray data can constrain the nature of the
inner accretion flow by measuring the morphology of the Fe K$\alpha$ line.
This emission line, with a rest energy of $6.4 \keV$ for neutral iron,
arises via fluorescence from the accretion
disk, which is illuminated by the Compton-scattered continuum X-rays.
Given the relatively high cosmic abundance
of iron and its high fluorescent yield, coupled with the lack of other lines
expected in that part of the spectrum, it is a reasonably ``clean'' probe of the
kinematics of the accreting material.

Narrow (usually unresolved by CCD detectors, i.e., $v/c \lesssim 0.005$) Fe K$\alpha$ lines have been
observed in the vast majority of Seyfert galaxies \markcite{Yaqoob2004}({Yaqoob} \& {Padmanabhan} 2004).  In
addition to their small width, the lack of variability implies that
they originate from the illumination by the primary X-ray
source of reprocessing material relatively far from the black hole, likely in
the outer disk or torus of Seyfert unification schemes
\markcite{Antonucci1993,Urry1995}({Antonucci} 1993; {Urry} \& {Padovani} 1995).  Indeed, this emission region has been
spatially resolved in the Seyfert 2 AGN NGC~4945
\markcite{Marinucci2012}({Marinucci} {et~al.} 2012), and it lies at a distance from the nucleus of $30-50 \pc$
($\sim10^8-10^9\,r_{\rm g}$ for the $10^6 \Msun$ black hole at its
core, where $r_{\rm g} \equiv GM/c^2$).  The near ubiquity of these features suggests
that this distant material is present in almost all Seyfert AGN.
In some Seyfert galaxies the Fe K line appears to
be broadened (to $v/c \gtrsim 0.1$), most likely by relativistic
effects; e.g., MCG--6-30-15, first observed by
\markcite{Tanaka1995}{Tanaka} {et~al.} (1995), and recently also confirmed in NGC~1365
\markcite{Risaliti2013}({Risaliti} {et~al.} 2013).  These sources are two of the best examples of
AGN displaying a prominent red wing indicative of
fluorescing material close to the innermost stable circular orbit (ISCO)
in the accretion disk.

A broad Fe K$\alpha$ line originating from material extending to the ISCO allows us to
determine whether the black hole is rotating, and if so, to
determine its spin and possibly direction as well (for recent reviews, see, e.g.,
\markcite{Reynolds2013}{Reynolds} 2013 and \markcite{Brenneman2013b}{Brenneman} 2013).
However, such broad, relativistic emission lines are not observed in all
Seyferts observed with high signal-to-noise (S/N)
\markcite{Nandra2007,dlCP2010,Brenneman2012}({Nandra} {et~al.} 2007; {de La Calle P{\'e}rez} {et~al.} 2010; {Brenneman} {et~al.} 2012), possibly indicating the absence of
relatively cold, Compton-thick gas close to the black hole (though the
caveats to line detection detailed in \markcite{Ballantyne2010}{Ballantyne} 2010 should also be kept in mind).  

Regardless of the mechanism by which they are determined, any inferences regarding the 
structure, location, and physical conditions of the accretion disk and the
corona require a precise, high S/N measurement of the
broad-band X-ray spectrum from $\leq 2$ to $\geq 50 \keV$.  This is
necessary in order to disentangle various emission
and absorption components contributing to the total observed X-ray
emission, described above.  A
significant advance towards such measurements is provided by the deployment of
the focusing hard X-ray telescopes onboard the 
{\it NuSTAR} observatory, the latest in the series of NASA's Small Explorer satellites.
This mission is sensitive in the bandpass of $3-79 \keV$ with the
updated calibration, and provides a
hundredfold improvement of sensitivity in the hard X-ray band over previous
instruments \markcite{Harrison2013}({Harrison} {et~al.} 2013).  The use of {\it NuSTAR} in
conjunction with X-ray telescopes that are more sensitive at softer
energies (e.g., {\it XMM-Newton} and {\it Suzaku}) yields the highest
S/N ever achieved across the $\sim0.2-79 \keV$ bandpass.

Equally important in deriving the physical properties of the disk and corona is
the selection of a representative, bright target.  One
good candidate is the southern Seyfert 1.2 galaxy IC~4329A ($z=0.0161$,
\markcite{Willmer1991}{Willmer} {et~al.} 1991; $N_{\rm H}\,[{\rm gal}]=4.61 \times 10^{20}
\pcmsq$, \markcite{Kalberla2005}{Kalberla} {et~al.} 2005; $M_{\rm BH}=1.20 \times 10^8 \Msun$,
\markcite{dlCP2010}{de La Calle P{\'e}rez} {et~al.} 2010), which in the hard
X-ray/soft $\gamma$-ray band appears similar to an average radio-quiet Seyfert
(e.g., \markcite{Zdziarski1996}{Zdziarski} {et~al.} 1996).  The host galaxy is an edge-on spiral
in a pair with IC~4329, separated by $\sim3$ arcmin.  IC~4329A was one of
the first AGN observed to have a Compton reflection component in addition to its strong
Fe K$\alpha$ line \markcite{Piro1990}({Piro}, {Yamauchi}, \& {Matsuoka} 1990).  As with most other X-ray emitting
Seyferts, it is variable, but the variability amplitude during a
typical observation is modest: the root mean square fractional
variability has been measured at $\leq20\%$ in the $15-150 \keV$ {\it RXTE} band
\markcite{Markowitz2009}({Markowitz} 2009), and $(17 \pm 3) \%$ in the $14-195 \keV$ band
with {\it Swift}/BAT \markcite{Soldi2014}({Soldi} {et~al.} 2014).  The average $2 - 10 \keV$ flux
historically ranges from $F_{\rm 2-10} \sim (0.1-1.8) \times 10^{-10} \ergpcmsqps$
\markcite{Beckmann2006,Verrecchia2007}({Beckmann} {et~al.} 2006; {Verrecchia} {et~al.} 2007).  IC~4329A has been the subject of many X-ray
observations, beginning with the analysis of its simultaneous {\it ROSAT} and {\it
  OSSE} spectrum \markcite{Zdziarski1994,Madejski1995}({Zdziarski} 1994; {Madejski} {et~al.} 1995).  
In harder X-rays, the source has also been observed by {\it BeppoSAX}
\markcite{Perola2002}({Perola} {et~al.} 2002), {\it ASCA+RXTE} \markcite{Done2000}({Done}, {Madejski}, \& {{\.Z}ycki} 2000) and {\it INTEGRAL} \markcite{Molina2013}({Molina} {et~al.} 2013), which have placed
rough constraints on the high-energy cutoff of the power-law (a proxy for
coronal temperature) at $E_{\rm cut} \geq 180 \keV$, $E_{\rm cut}=150-390 \keV$ and $E_{\rm
  cut}=60-300 \keV$, respectively.  Combining the non-simultaneous {\it INTEGRAL} and {\it
  XMM-Newton} data further constrained the cutoff energy to $E_{\rm
  cut}=130-203 \keV$ \markcite{Molina2009}({Molina} {et~al.} 2009), while a combination of the {\it
  XMM} and {\it BeppoSAX} data yielded $E_{\rm cut}=150-390 \keV$ \markcite{Gondoin2001}({Gondoin} {et~al.} 2001).

A detailed examination of the {\it ASCA} and simultaneous {\it RXTE} data revealed that the
continuum is indeed described well by the model used to describe the {\it
  ROSAT+OSSE} data (either thermal or non-thermal Comptonization plus
neutral, distant reflection), and that the Fe K$\alpha$ line is moderately broadened and can be described
by a Gaussian with FWHM of $\sim30,000 \kmps$ \markcite{Done2000}({Done} {et~al.} 2000).  This
is consistent with the conclusions of \markcite{Dadina2007}{Dadina} (2007), who 
noted a moderately broad Fe K$\alpha$ line of similar width and equivalent widths up
to $EW \sim 180 \eV$ in {\it BeppoSAX} data, paired with measured reflection fractions up to
$R \sim 1.5$.  Both the {\it
  ASCA} and {\it ROSAT} data, as well as the {\it XMM-Newton}
observations \markcite{Steenbrugge2005}({Steenbrugge} {et~al.} 2005), suggest that the soft
X-ray spectrum is absorbed by a combination of neutral and partially ionized
gas, with a total column of $\sim3 \times 10^{21} \pcmsq$.  This is
comparable to the host galaxy's ISM column density
\markcite{Wilson1979}({Wilson} \& {Penston} 1979).  After accounting
for the reflection component, the source shows some modest spectral variability
of the primary continuum, being softer at higher flux levels
\markcite{Madejski2001,Miyazawa2009,Markowitz2009}({Madejski}, {Done}, \&  {{\.Z}ycki} 2001; {Miyazawa}, {Haba}, \&  {Kunieda} 2009; {Markowitz} 2009).

Here, we report on results from our simultaneous {\it Suzaku} and {\it NuSTAR} observation of
IC~4329A.  We discussed our measurements of the properties of the
underlying continuum in \markcite{Brenneman2014}{Brenneman} {et~al.} (2014) (hereafter referred to
as paper I), and in this work we update those values and focus on constraining the reprocessing
components.  In \S2, we report on the {\it Suzaku} and {\it NuSTAR} observations, and in
\S3 we present a brief timing analysis of the data.  Our spectral analysis
follows in \S4, with a discussion of the inferred accretion disk properties and their
implications in \S5.

\section{Observations and Data Reduction}
\label{sec:data}

IC~4329A was observed quasi-continuously and contemporaneously by 
{\it Suzaku} and {\it NuSTAR} from
August 12-16, 2012 in normal clocking mode.  The observations had a
roughly $50\%$ efficiency due to Earth occultations.  After eliminating Earth
occultations, passages through the South Atlantic Anomaly (SAA) and other periods
of high background, the {\it Suzaku} observation
totaled $\sim118 \ks$ of on-source time from August 13-15, while the {\it NuSTAR} observation
totaled $\sim160 \ks$ of on-source time from August 12-16.  Count rates, total
counts and signal-to-noise (S/N) ratio for each instrument from the two
observatories is listed in Table~\ref{tab:obs}.

\begin{table}
\begin{center}
\begin{tabular}{|ccccc|} 
\hline\hline
{\bf Instrument} & {\bf Exposure (ks)} & {\bf Count Rate (cts/s)} & {\bf Total
  Counts} & {\bf S/N} \\
\hline
{\it Suzaku}/XIS~0 & $118$ & $3.792 \pm 0.006$ & $448,350$ & $629$ \\
{\it Suzaku}/XIS~1 & $118$ & $3.732 \pm 0.006$ & $442,833$ & $652$ \\
{\it Suzaku}/XIS~3 & $118$ & $3.980 \pm 0.006$ & $470,300$ & $644$ \\
{\it Suzaku}/PIN   & $106$ & $0.268 \pm 0.003$ & $53,426$  & $5$ \\
{\it NuSTAR}/FPMA  & $162$ & $2.608 \pm 0.004$ & $426,274$ & $653$ \\
{\it NuSTAR}/FPMB  & $159$ & $2.513 \pm 0.004$ & $403,588$ & $645$ \\
\hline\hline
\end{tabular}
\end{center}
\caption{\small{Observation details for the {\it Suzaku} and {\it NuSTAR}
    campaign on IC~4329A. Exposure times, count rates, total counts and S/N are
    taken from the background-subtracted data.  Energy ranges for the XIS
    detectors are $0.7-1.5$ and $2.5-10 \keV$, $16-60 \keV$ for the PIN, and
    $3-79 \keV$ for the FPMA and FPMB data.  Signal-to-noise was
    calculated for the unbinned spectra over the given energy bands using the source-dominated
    case described in \markcite{Longair2011}{Longair} (2011) for the XIS, FPMA and FPMB.
    The PIN data were background-dominated and thus rebinned to
    $S/N=5$.}}
\label{tab:obs}
\end{table}

The {\it Suzaku}/XIS data were taken with the telescope at the XIS nominal
aimpoint, and were reduced as per the ABC
Guide\footnote{http://heasarc.gsfc.nasa.gov/docs/suzaku/analysis/abc/}, using
the latest versions of the CALDB (October 2013) and HEASoft (v6.15) packages as of the time of
this writing.  After reprocessing the data from XIS~0, XIS~1
and XIS~3 (XIS~2 has been inoperable since November 2006) using the {\tt aepipeline}
script, source and background regions were extracted for each detector within
{\sc xselect}.  Source regions were circular and 160 arcseconds in radius
centered on the source, while background regions were extracted from as much of
the surrounding region on the same chip as possible, avoiding the source region
and the calibration sources in the detector corners.  We then generated response
matrices using the {\tt xisresp} script at ``medium'' speed, after which we
combined the data from the front-illuminated (FI) XIS~0 and XIS~3 detectors using the
{\tt addascaspec} script in order to maximize S/N.  The XIS source and
  background spectra, as well as the responses, were then grouped to a minimum
  of 25 counts per channel in order to facilitate robust $\chi^2$ fitting.  For
  all of the spectral fitting presented later in \S\ref{sec:spectral},
  we evaluate the combined XIS-FI with the XIS~1 spectra between
  $0.7-1.5$ and $2.5-10 \keV$.  The $1.5-2.5 \keV$ range is ignored
  due to the presence of calibration features.  We allow for a 
global flux cross-normalization error between the XIS-FI and XIS~1
spectra, fitting it as a free parameter.  At $0.975 \pm 0.005$, we
find it slightly lower than the published
value of the XIS~1 cross-normalization relative to the
combined XIS-FI data ($1.019
\pm 0.010$).\footnote{http://heasarc.gsfc.nasa.gov/docs/suzaku/analysis/watchout.html}

The HXD/GSO detection of IC~4329A was marginal,  
but corresponds, very roughly, to $F_{\rm 50-150} \sim 10^{-10} \ergpcmsqps$ in
the $50-150 \keV$ band.  With
such a weak detection, we did not use the GSO data in our analysis.

Data from the HXD/PIN instrument were again reduced and
reprocessed as per the {\it Suzaku} ABC Guide.
For background subtraction, we used the ``tuned'' non X-ray background
(NXB) event file for August 2012 from the {\it Suzaku} CALDB, along with
the appropriate response file and flat field file for epoch 11 data.  We modeled
the cosmic X-ray background (CXB) contribution as per the
ABC Guide, simulating its spectrum in {\sc xspec} \markcite{Arnaud1996}({Arnaud} 1996).
The simulated CXB spectrum (modeled as in \markcite{Boldt1987}{Boldt} 1987) contributed a count rate of $(14.7 \pm
0.1) \times 10^{-3} \ctsps$ to the total X-ray background from $16-60 \keV$, coupled
with a source count rate of $(268 \pm 3) \times 10^{-3} \ctsps$ over this range.  
The NXB and CXB files were combined to form a single PIN background spectrum.
Because the PIN data only contain 256 spectral channels (vs. the 4096
channels in the unbinned XIS data), rebinning to 25 counts per bin was
not necessary in
order to facilitate $\chi^2$ fitting.  Rather, we rebinned the PIN
spectrum to have a S/N of 5 in each energy bin, which limited our energy
range to $16-60 \keV$.  We also added $3\%$ systematic errors to the PIN
data to account for the uncertainty in the non-X-ray background data
supplied by the {\it Suzaku} calibration team.  
For most of the spectral fitting presented in this paper,
we assume a PIN/XIS-FI cross-normalization factor of $1.164 \pm 0.014$ as per the 
{\it Suzaku} memo
2008-06\footnote{http://heasarc.gsfc.nasa.gov/docs/suzaku/analysis/watchout.html},
though we allow the factor to vary around this value.  The fitted
value is $1.217 \pm 0.024$.

The {\it NuSTAR} data were collected with the two focal plane module
telescopes (FPMA and FPMB) centered roughly $\sim2$ arcmin from the
nucleus of IC~4329A.  We reduced the data using the {\it NuSTAR} Data
Analysis Software ({\sc nustardas}) and calibration version
1.1.1\footnote{http://heasarc.gsfc.nasa.gov/docs/nustar/}.  We filtered the event
files and applied the default depth
correction using the {\tt nupipeline} task.  The
source and background regions were circles of
radius 75 arcsec, with the source region centered on IC~4329A and the
background region taken from the corner of the same detector, as close
as possible to the source without being contaminated by the PSF wings.
Spectra
and light curves were extracted and response files were generated using the {\tt
  nuproducts} task.  In order to minimize systematic effects, we have
not combined responses or spectra from FPMA and FPMB; instead, we fit
them simultaneously.  We allow the cross-normalization factor
between each module and the {\it Suzaku}/XIS-FI data to fit freely.
The absolute cross-calibration factor is $1.092 \pm 0.018$ for FPMA
and $1.166 \pm 0.020$ for FPMB.  We obtain excellent agreement in
the expected spectral shape below $10 \keV$, where there is good overlap
between {\it Suzaku}/XIS and {\it NuSTAR}/FPMA and FPMB: when fit with
a simple power-law model, the photon index for the {\it NuSTAR} detectors is
$\Gamma=1.687 \pm 0.002$, while that of {\it Suzaku}/XIS is $\Gamma=1.690
\pm 0.003$.

For all the analysis presented here, we used {\sc xronos} version 5.22 and
{\sc xspec} version 12.8.1, along with other {\sc ftools} packages within
HEASoft 6.14.  Uncertainties quoted within the text are at the 1$\sigma$ level of
confidence, unless otherwise specified, while those in tables are quoted at
$90\%$ confidence.

\section{Timing Analysis}
\label{sec:timing}

\begin{figure}
\hbox{
\includegraphics[width=0.6\textwidth,angle=270]{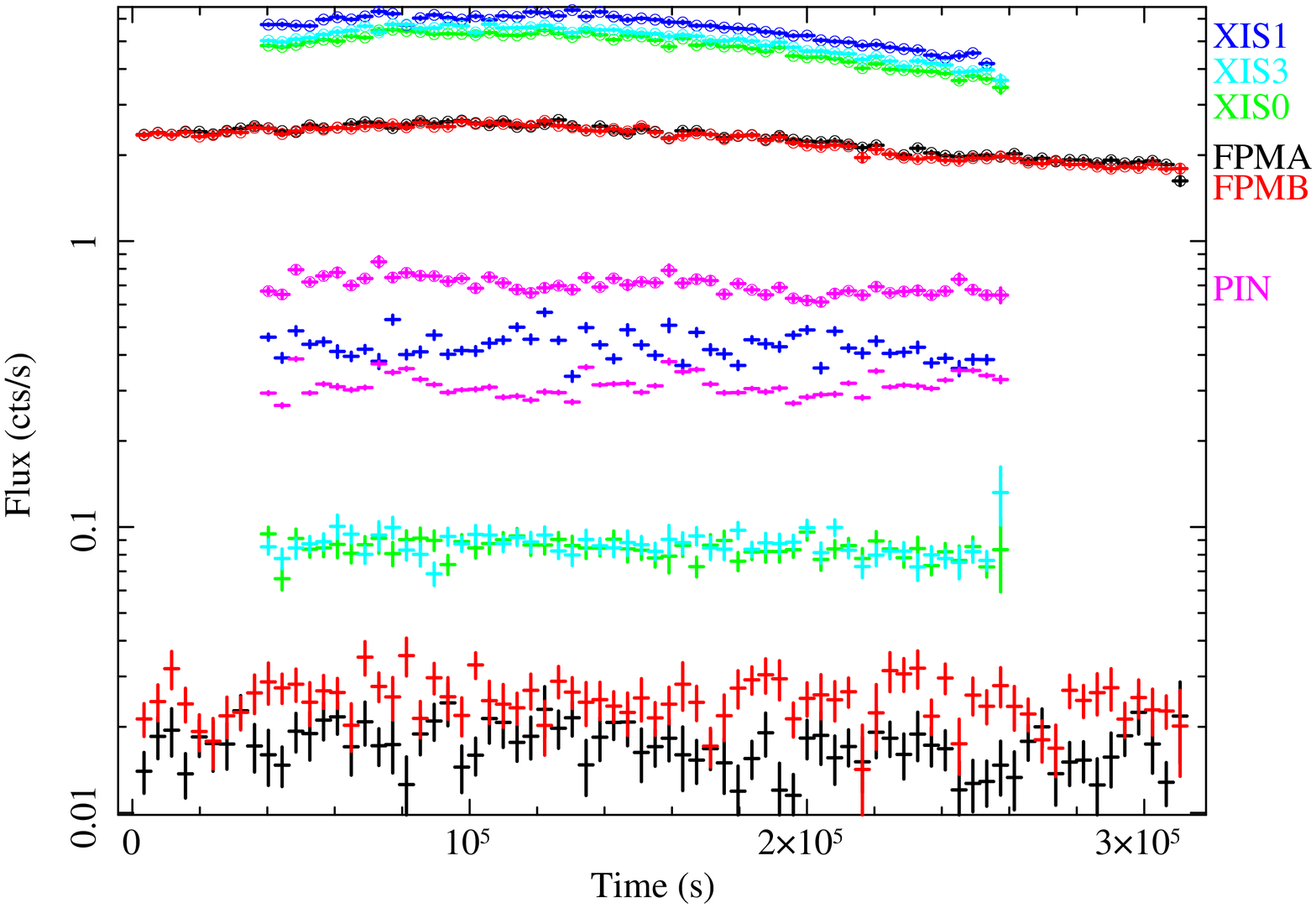}
}
\caption{{\small Simultaneous source and background light curves from the six instruments on
{\it Suzaku} and {\it NuSTAR} that observed IC~4329A simultaneously in
2012, integrated over all energies for each detector.  Source counts
are shown with crosses in circles, background counts with plain crosses.
Black data points show the {\it NuSTAR}/FPMA data, red show the FPMB
data, green show the {\it Suzaku}/XIS~0 data, dark blue 
show the XIS~1 data, light blue show the XIS~3 data, and magenta show the
HXD/PIN data.  Background colors
match the source colors and the vertical order of appearance is the same as for
the source data points, except for the PIN background, which is second from the
top of the background data points.}}
\label{fig:src_bg_lcs}
\end{figure}

IC~4329A showed a modest, secular flux evolution during the
joint {\it Suzaku/NuSTAR} observing campaign, roughly consistent with previous
observations \markcite{Markowitz2004}({Markowitz} \& {Edelson} 2004).  In both datasets, the
source flux increased by $\sim12\%$ over the first $50 \ks$ of the
exposure, then plateaued at the maximum count rate for $\sim50 \ks$
before decreasing by $\sim34\%$ over the remainder of the
observation (clock time is used to measure these intervals).  No
significant flux variability was seen on short
timescales in any of the instruments.  Background levels were
approximately constant, except for a factor $\sim3$ flare in {\it
  Suzaku}/XIS~1 seen in the last time bin (see
Fig.~\ref{fig:src_bg_lcs}).  On average, the $2-10 \keV$ flux during our
observations was comfortably within the historical
range\footnote{http://ned.ipac.caltech.edu/}: $1.02 \times
10^{-10} \ergpcmsqps$ vs. $(0.1-1.8) \times 10^{-10}
\ergpcmsqps$.  

\begin{figure}
\centerline{
\includegraphics[width=0.35\textwidth,angle=270]{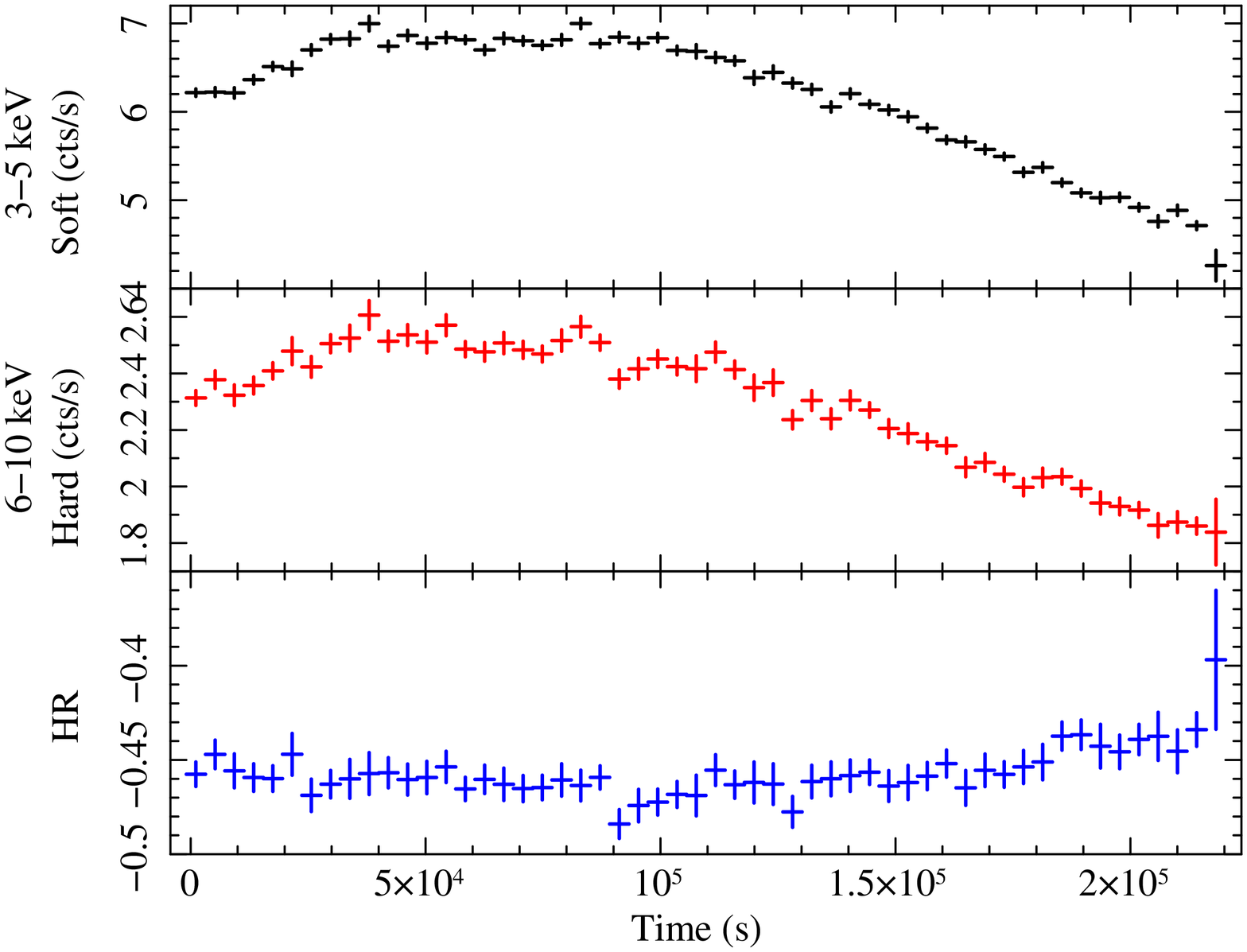}
}
\centerline{
\includegraphics[width=0.35\textwidth,angle=270]{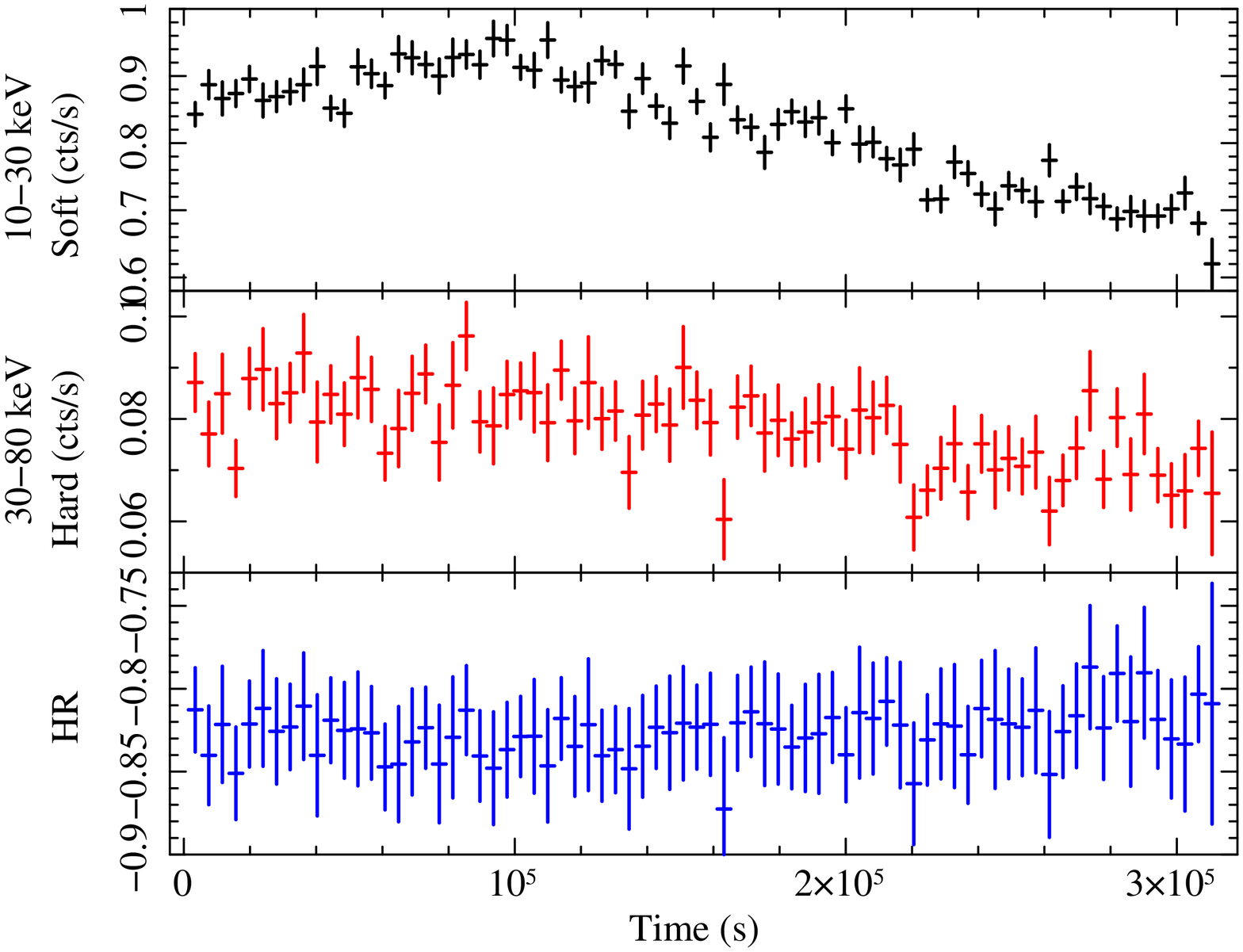}
}
\centerline{
\includegraphics[width=0.35\textwidth,angle=270]{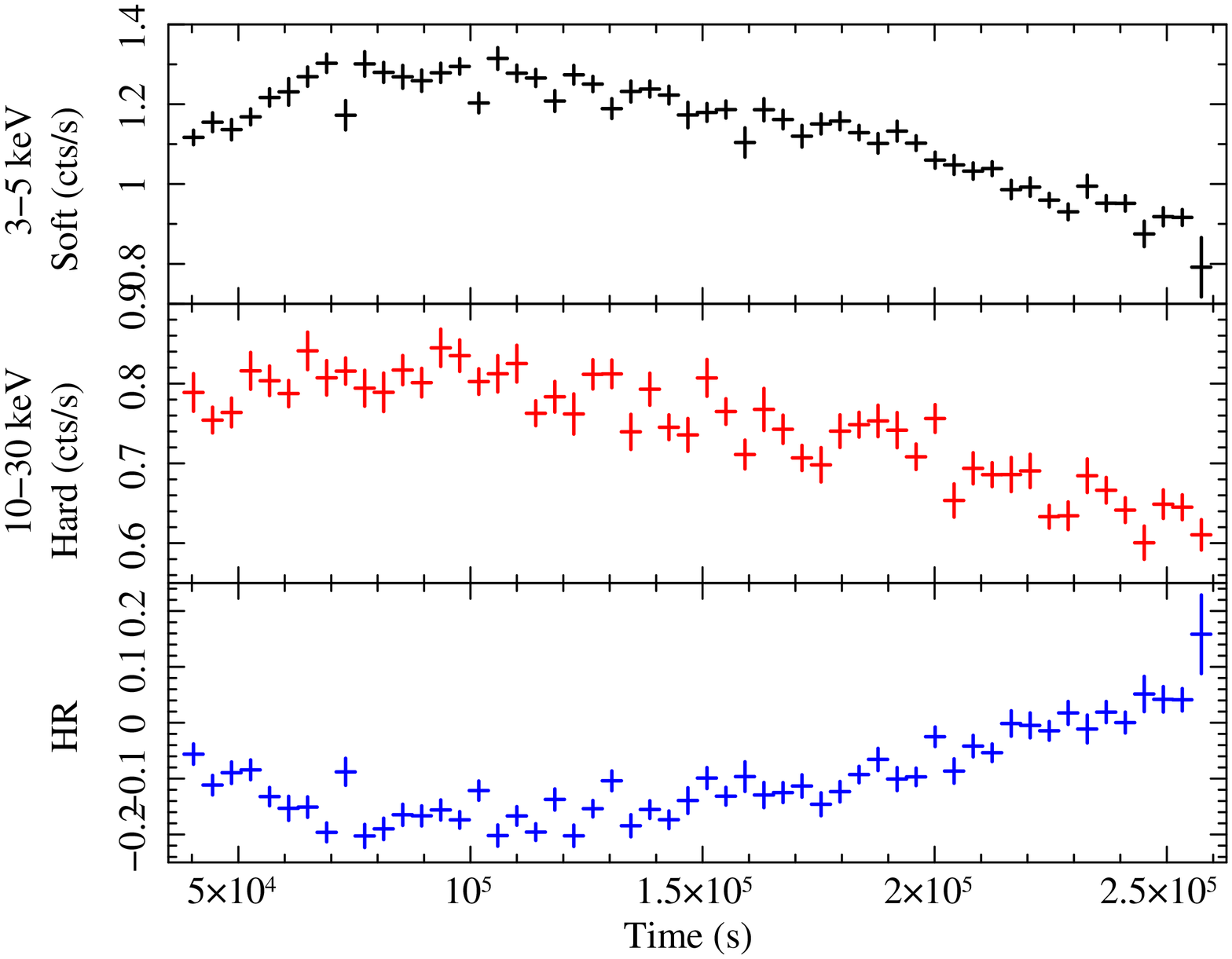}
}
\caption{{\small {\it Top:} Simultaneous light curves and hardness ratio for the combined
    {\it Suzaku}/XIS data at a resolution of 4096 sec/bin.  The top panel shows the $3-5 \keV$ data, middle
    shows the $6-10 \keV$ data and bottom shows the hardness ratio between them,
computed as (H-S)/(H+S), where $H$ and $S$ denote the hard and soft
count rates, respectively.  {\it Middle:} The same plot, this time for the
{\it NuSTAR}/FPMA data using $10-30 \keV$ for the soft data and $30-79 \keV$ for the
hard data.  The FPMB light curves and hardness ratio are virtually identical to
those from FPMA, so are not shown here.  {\it Bottom:} The same plot,
this time comparing the $10-30 \keV$ {\it NuSTAR}/FPMA light curve to the
$3-5 \keV$ light curve taken from {\it Suzaku}/XIS~3.  The light
curves are required to be strictly simultaneous and are normalized by
their effective areas in the given energy ranges in order to account
for the differences between instruments.}}
\label{fig:lchr}
\end{figure}

The hardness ratio in the XIS was nearly constant over the observation
(Fig.~\ref{fig:lchr}), taken between the $2-5
\keV$ and $5-10 \keV$ bands.  During the final $\sim50 \ks$ the hardness did
increase by $\sim13\%$ as the source continued to decrease in flux by
$\sim19\%$ over the same time interval.  The {\it NuSTAR} data show a
similar overall trend, visually, but the total amount of flux
variability is significantly lower at these higher energies
($10-30 \keV$ vs. $30-79 \keV$) as compared with the {\it Suzaku}/XIS
data.  The hardness ratio taken from the {\it NuSTAR}
data is consistent with a constant value (Fig.~\ref{fig:lchr}).

\begin{figure}
\hbox{
\includegraphics[width=0.6\textwidth,angle=270]{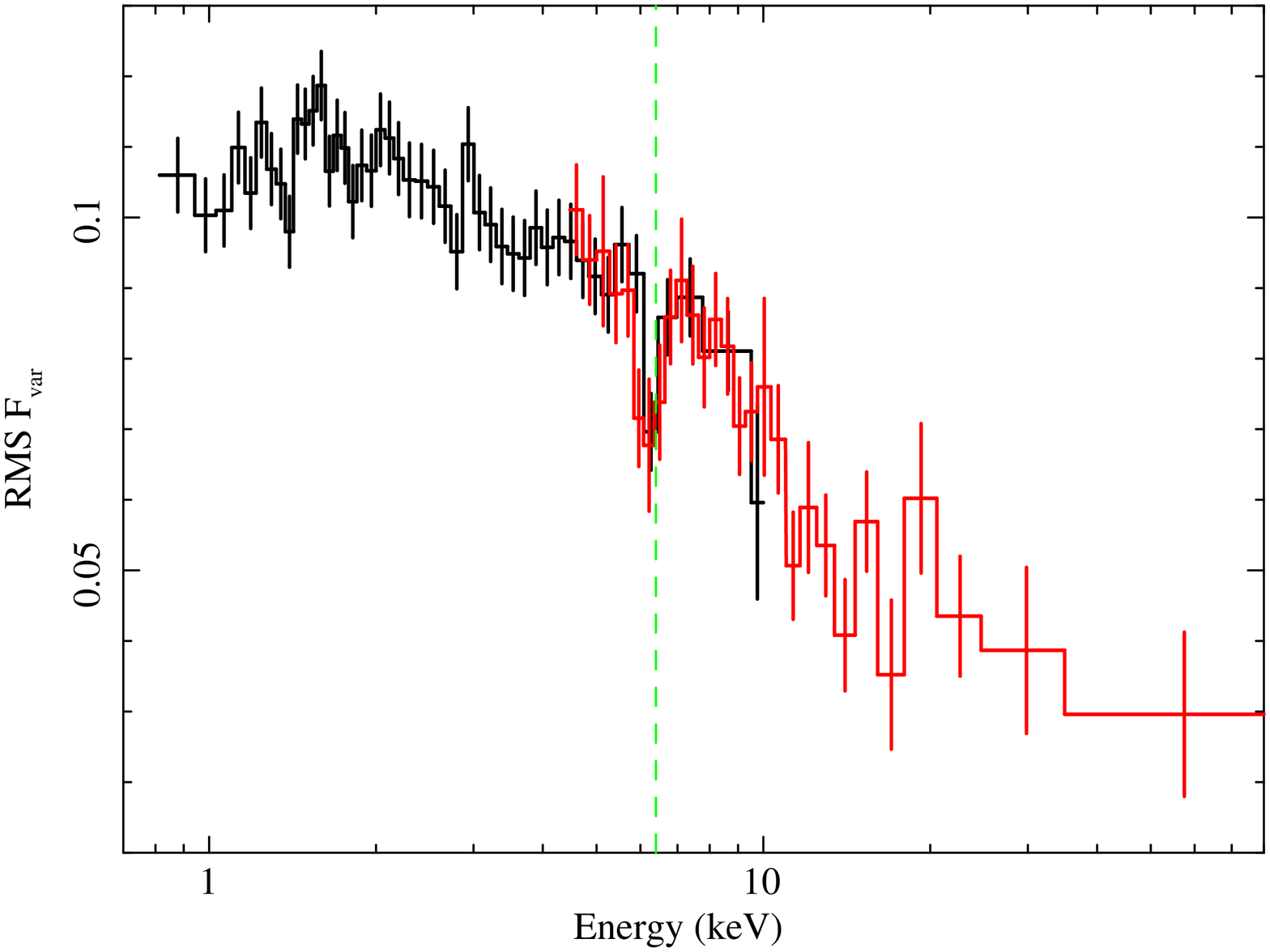}
}
\caption{{\small Root mean square fractional variability vs. energy for the {\it
  Suzaku}/XIS-FI (black) and {\it NuSTAR}/FPMA (red)
    data (FPMB is omitted for clarity, but overlaps with FPMA).  Note
  the prominent, sharp dip in RMS $F_{\rm var}$ at $\sim6.4 \keV$,
    the rest-frame energy of the Fe K$\alpha$ line (this energy is
    marked with the dashed green line).  This dip indicates that the
narrow Fe K$\alpha$ line is less variable than the surrounding continuum, as one
would expect if it originates from gas that is distant from the source of the
hard X-rays.}}
\label{fig:fvar}
\end{figure}

To assess the amount of variability in our observations in a
model-independent way, we also calculated the root mean square
fractional variability vs. energy of the combined {\it Suzaku}/XIS data as well
as the {\it NuSTAR}/FPMA and FPMB data.  This fractional variability
spectrum, or RMS F$_{\rm var}$, is derived using the methods described
in \markcite{Edelson2002}{Edelson} {et~al.} (2002) and \markcite{Vaughan2003}{Vaughan} {et~al.} (2003), and the results are plotted in
Fig.~\ref{fig:fvar}.  The source shows the expected decrease in overall fractional source
variability with energy due to a combination of the increasing relative importance of
reflection from distant matter \markcite{Niedzwiecki2010}({Nied{\'z}wiecki} \& {Miyakawa} 2010) as well as the
decreasing importance of the power-law component.  We also note that both the
{\it Suzaku} and {\it NuSTAR} data display a prominent dip between
$6-7 \keV$, roughly coincident with the Fe K$\alpha$ line thought to
arise from the fluorescence of distant material due to irradiation by the primary X-ray
source.  If this emitting gas is located at many thousands of
gravitational radii from the corona, as described in AGN unification
schemes, it would not vary on the
timescales of our observing campaign, meaning that we should expect to
see such a dip in the RMS F$_{\rm var}$ spectrum.  

Given the lack of short timescale variability in our
observations of IC~4329A, along with its overall modest flux and
hardness ratio changes,
we use the time-averaged spectrum
in all of the broad band spectral fitting
(\S\ref{sec:suzaku}-\ref{sec:broadband}).   
A discussion of the modest
spectral variability is deferred to \S\ref{sec:spec_var}.  

\section{Spectral Analysis}
\label{sec:spectral}

We begin our analysis of the time-averaged spectrum of IC~4329A with
an examination of the {\it Suzaku}/XIS spectrum from $0.7-10.0 \keV$ (ignoring
the energy range from $1.5-2.5 \keV$ due to calibration uncertainties around the
Si edge).
We constrain the properties of the continuum and get a preliminary
assessment of the contributions of complex absorption intrinsic to the
AGN, distant reflection from the outer disk and/or torus, and
relativistic inner disk reflection.  We then add in the {\it
  Suzaku}/PIN data from $16-60 \keV$ to aid in constraining the slope of the power-law
continuum and the fractional contribution of reflection.  

We continue our analysis by
including the {\it NuSTAR}/FPMA and FPMB data, which have significantly
better signal-to-noise and reduced systematic error than the PIN above
$10 \keV$, and to access energies out to $79 \keV$ that are
unreachable by the {\it Suzaku}/PIN instrument.  The use of
simultaneous {\it Suzaku} and
{\it NuSTAR} data allows us to better constrain the parameters of the
continuum and reflection, and to constrain the cutoff energy of the
power-law component.  The importance of having broadband, high S/N
X-ray spectra from both observatories lies in our enhanced ability to
break modeling degeneracies, and to therefore constrain the physical
parameters of three main spectral components with enough accuracy and precision to yield the
best estimates of the temperature and optical depth of the coronal
plasma taken to date.  

We also consider the spectral variability of the source, examining the
high-flux vs. low-flux spectra in order to understand the physical
processes driving the flux evolution in IC~4329A.

\subsection{A First Look at the {\em Suzaku} Spectra}
\label{sec:suzaku}

The {\it Suzaku}/XIS-FI and XIS~1 data between $2.5-5 \keV$ and
$7.5-10 \keV$ are well fit by a power-law of slope $\Gamma=1.66 \pm 0.01$ modified by a Galactic
column of $N_{\rm H}=4.61 \times 10^{20} \pcmsq$
\markcite{Kalberla2005}({Kalberla} {et~al.} 2005).  Here we use the {\tt TBabs}
model of \markcite{Wilms2000}{Wilms}, {Allen}, \& {McCray} (2000), along with abundances set to {\tt wilm} \markcite{Wilms2000}({Wilms} {et~al.} 2000)
and cross-section table set to {\tt vern} \markcite{Verner1996}({Verner} {et~al.} 1996).  Strong
residuals remain when the entire bandpass is considered, however (the XIS~1 data are ignored above $7 \keV$ due to a
rapid loss in detector sensitivity above this energy).  
Below $2 \keV$, these residuals take the form of a
pronounced dip below the ideal data/model ratio of unity, indicating
the presence of a significant column of absorbing gas along the line
of sight to the AGN.  The initial power-law model has a global
goodness-of-fit of $\chi^2/\nu=444071/2147\,(207)$.  The spectral fit improves
dramatically with the addition of an {\sc xstar} \markcite{Kallman2001}({Kallman} \& {Bautista} 2001) warm
absorber table in which the absorbing gas has
a column density of $N_{\rm H} = (5.9 \pm 0.1) \times 10^{21} \pcmsq$ and an
ionization parameter of ${\rm log} \, \xi = 0.80 \pm 0.01 \ergcmps$:
this results in $\chi^2/\nu=3558/2145\,(1.66)$.  

\begin{figure}
\centerline{
\includegraphics[width=0.35\textwidth,angle=270]{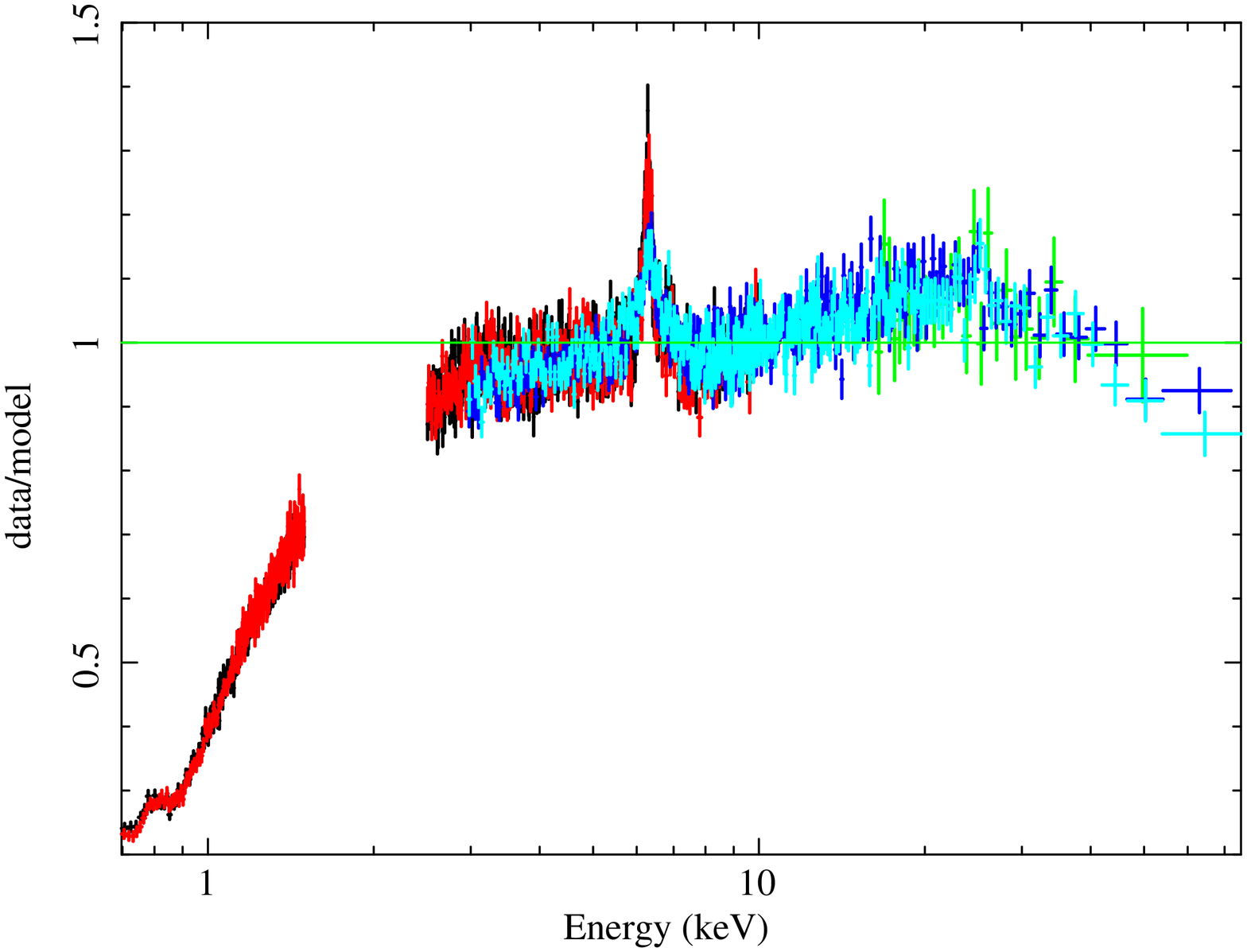}
}
\centerline{
\includegraphics[width=0.35\textwidth,angle=270]{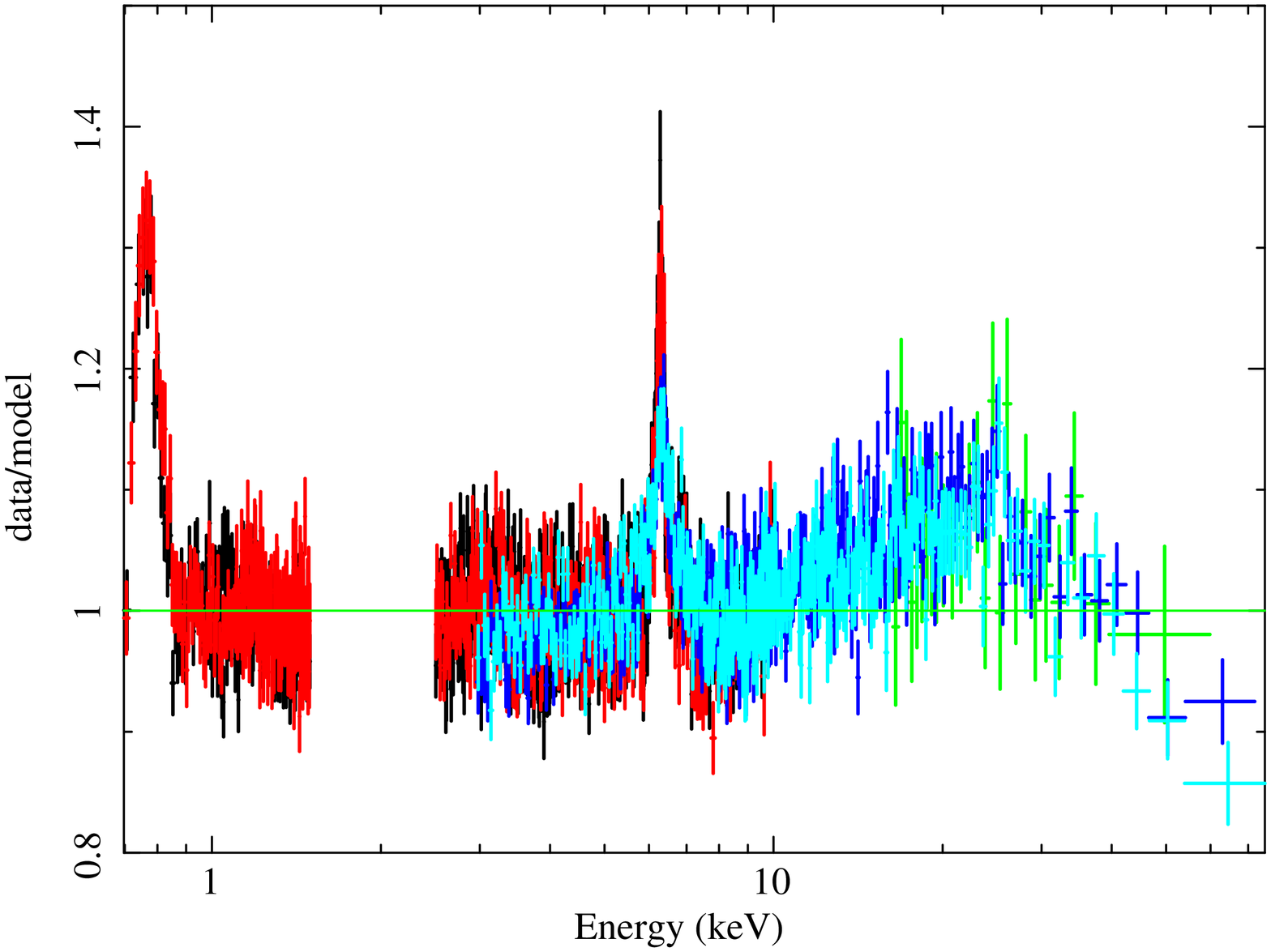}
}
\centerline{
\includegraphics[width=0.35\textwidth,angle=270]{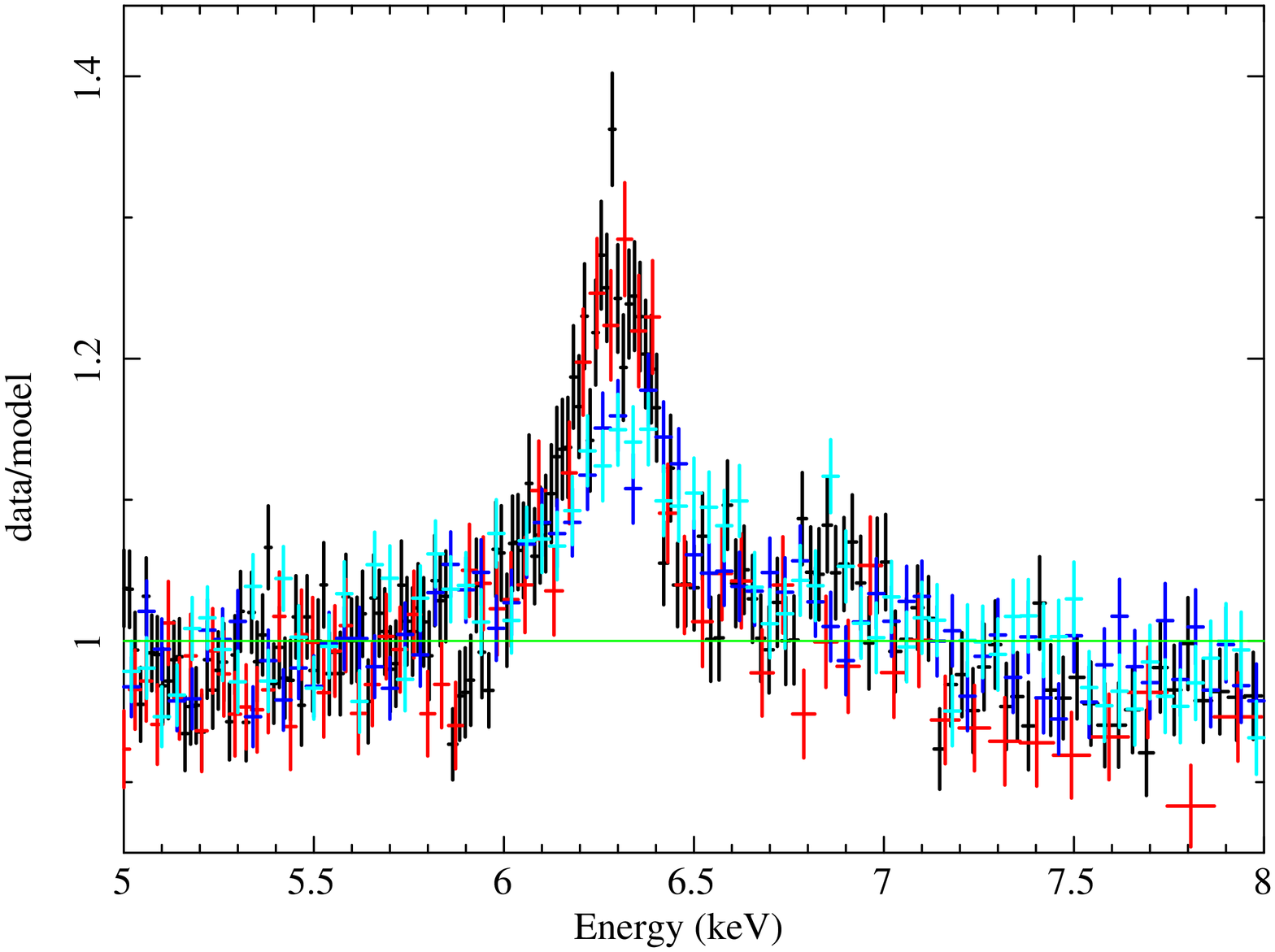}
}
\caption{{\small {\it Top:} The ratio of the spectrum of IC~4329A to a power-law
    modified by Galactic photoabsorption.  Co-added {\it Suzaku}/XIS-FI data are
    in black, XIS~1 data are in red, HXD/PIN data are in green, {\it
      NuSTAR}/FPMA data are in dark blue, and FPMB data are in light blue.  The
    horizontal green line represents a perfect fit, or a data-to-model ratio of
    unity.  {\it Middle:} Same plot as above, but absorption
    intrinsic to the AGN is now included in the model, eliminating the
    curvature below $2 \keV$.  {\it Bottom:}
    A zoomed-in version of the middle plot, focused on
    the Fe K region.}}
\label{fig:phpo_rat}
\end{figure}

An emission line-like feature remains in the residuals at $0.78 \pm 0.01
\keV$.  This feature can be modeled with a Gaussian that likely
represents a blend of the resonance, intercombination and forbidden
O\,{\sc vii} emission lines.  The
spectral resolution of the {\it Suzaku}/XIS instrument is insufficient
to separate these three putative lines, however, so it is not possible
to derive a density or temperature for the emitting gas via line
ratios.  This residual feature also does not appear to be a
simple consequence of using only one layer of intervening gas to
describe the intrinsic absorption of the source, since adding in a
second absorber does not improve the fit, nor does it mitigate this
feature.  If modeled with a Gaussian, the line has a width
of $\sigma=0.02 \pm 0.01 \keV$ (FWHM$\sim 18,000 \pm 9000 \kmps$; well within the Broad
Emission Line Region, though if this is a line blend any
velocity inferred from its width would be erroneous) and a strength relative to the continuum of
$EW=39 \pm 6 \eV$.  Adding in this feature
further improves the fit to $\chi^2/\nu=3151/2142\,(1.47)$.  The
succession of data-to-model ratios from the
simple power-law and absorbed power-law models is shown in
Fig.~\ref{fig:phpo_rat}, which also highlights the putative
O\,{\sc vii} line and the residuals in the Fe K band.  

Above $5 \keV$, the Fe K emission complex is
clearly present, with a prominent, narrow Fe K$\alpha$ line at $6.41
\pm 0.04
\keV$ and perhaps a blend of Fe K$\beta$ and Fe\,{\sc xxvi} at $6.99
\pm 0.03 \keV$ (Fig.~\ref{fig:phpo_rat}).  Though there is a visual suggestion of an Fe\,{\sc
  xxv} emission line at $\sim6.7 \keV$, including a Gaussian at this
energy does not improve the fit.  The iron edge is also clearly seen at $7.25 \pm
0.07 \keV$, perhaps indicating a slightly ionized disk, consistent
with the possible presence of an ionized iron line at $\sim7 \keV$.  No
strong, broadened emission in the Fe K band is immediately apparent to the
eye, though there are indications of a possible Compton shoulder or
iron line red wing at
$6.2 \pm 0.04 \keV$.  If we use Gaussian emission lines and an absorption edge to
parametrize these features and require that all the Fe K emission
features have the same width as the narrow Fe K$\alpha$ line, we
obtain a width of $\sigma=0.05 \pm 0.01 \keV$, $EW_{\rm K\alpha}=64 \pm 6 \eV$,
$EW_{\rm FeXXVI/K\beta}=12 \pm 3 \keV$, and $EW_{\rm CS}=13 \pm 4 \eV$.  The
goodness-of-fit for the $0.7-10 \keV$ XIS spectra including these
components is
$\chi^2/\nu=2162/2133\,(1.01)$.  

Substituting a broad Gaussian emission line at
$6.4 \keV$ in place of the Compton shoulder results in the same
overall goodness-of-fit, with the width of the broad line at
$\sigma=0.33 \pm 0.09 \keV$.  This corresponds to $FWHM \sim 36,000 \pm
9800 \kmps$, placing the
origin of the line at $r \sim 70 \pm 19 \,r_{\rm g}$, if
the width of the line corresponds to a Keplerian velocity in the disk.
This velocity is consistent with that found by \markcite{Shu2010}{Shu}, {Yaqoob}, \& {Wang} (2010) using {\it
  Chandra}/HETG data, and is $\sim5$ times greater than that of the
H$\beta$ line in IC~4329A ($FWHM=5620 \pm 200 \kmps$; \markcite{Shu2010}{Shu} {et~al.} 2010 and
references therein).  This implies that the X-ray broad line emission
region (BELR) is $\sim5$ times closer to the black hole than the
optical BELR.
The inclusion of a relativistic line via the {\tt
  relline} model \markcite{Dauser2010}({Dauser} {et~al.} 2010) does not improve the fit, however,
and the {\tt relline} parameters are
unconstrained except for the normalization.  We note that the joint {\it
  ASCA} and {\it RXTE} analysis of IC~4329A by \markcite{Done2000}{Done} {et~al.} (2000) included
a similar broadening of the Fe K$\alpha$ line, but likewise could not
definitively conclude an inner disk origin for this feature.  We proceed under the assumption
that the excess emission redward of $6.4 \keV$ is either a
Compton shoulder or a modest contribution from inner disk
reflection.  Its equivalent width is $EW=36 \pm 8 \eV$ in the case of a
broad Gaussian at $6.38 \pm 0.02 \keV$.

Adding in the HXD/PIN data from $16-60 \keV$, we notice convex
curvature that peaks around $\sim25 \keV$ and tails off at higher
energies, suggesting the presence of the Compton reflection continuum
and a high-energy cutoff to the power-law component, as found in 
paper I.  
The addition
of the unmodeled PIN data results in a predictable
worsening of the goodness-of-fit to $\chi^2/\nu=2267/2201\,(1.03)$
before refitting, and $\chi^2/\nu=2248/2201\,(1.02)$ after refitting.  Including a
{\tt pexrav} component \markcite{Magdziarz1995}({Magdziarz} \&
{Zdziarski} 1995) with $R=0.19 \pm 0.05$ recovered the $\leq10
\keV$ fit of $\chi^2/\nu=2230/2203\,(1.01)$.  We note that this
reflection includes both a contribution from the neutral outer disk or
torus, as well as the inner disk (if present).  The inclination angle was unconstrained in the fit, so we fixed it to
$i=60 \degmark$; this is the typical value assumed for the average
over a distribution of disks at random angles.  We elected to keep the cutoff energy
of the illuminating power-law fixed at $300 \keV$, since the limited energy range and high background of
the PIN data render them of limited use in probing this parameter.  

For a more self-consistent approach, we then replaced
the Gaussians-plus-{\tt pexrav} model with a {\tt pexmon} component
\markcite{Nandra2007}({Nandra} {et~al.} 2007), which calculates the expected Fe K emission
signatures (Fe K$\alpha$, K$\beta$ and the Compton shoulder, as well
as Ni K$\alpha$ and K$\beta$) and the
corresponding Compton hump together.  We fixed the cutoff energy
of the power-law for the {\tt pexmon} model at $300 \keV$.  Allowing the iron abundance to
fit freely, we obtained $R=0.21 \pm 0.04$ and
Fe/solar$=1.56 \pm 0.64$, for a goodness-of-fit of
$\chi^2/\nu=2229/2203\,(1.01)$.  The inclination angle was fixed to
$i=60\degmark$ as in the {\tt pexrav} fit.
Including Gaussian emission lines for Fe\,{\sc xxv} and {\sc xxvi} did
not statistically improve the fit, though they did lessen the
residuals in these regions to the eye.  A visual inspection of the
residuals also indicated that the region from $6.0-6.4 \keV$ was underfitted by the
model.  Given that {\tt pexmon} includes a Compton shoulder already,
we infer that this excess emission likely corresponds to a modest broad Fe K
line.  Parametrizing this component with a Gaussian emission line at $6.36 \pm 0.03 \keV$
yields an equivalent width of $EW=34 \pm 7 \eV$, consistent
with its value above, and also shows a similar line width.

%
Replacing the {\tt pexmon}
component with the more physically realistic {\sc xstar}-generated {\tt xillver} model of
\markcite{Garcia2014}{Garc{\'{\i}}a} {et~al.} (2014) yields a
similar goodness-of-fit, with $\chi^2/\nu=2227/2200\,(1.01)$ for
Fe/solar$=1.3 \pm 0.5$ and $K_{\rm refl}=(4.50 \pm 1.38) \times 10^{-5}
\phpcmsqps$ (roughly equivalent to $R=0.20$ in {\tt pexmon}).  The
ionization we find is low but unconstrained, so we elected
to fix it at the neutral value.  The inclination angle of the
reprocessor to the line of sight was similarly unconstrained, so we
fixed it to $i=60 \degmark$.  As with
the {\tt pexmon} model, {\tt xillver} underestimated the amount of
emission between $6.0-6.4 \keV$ in spite of having the Compton
shoulder included in the model; including an additional Gaussian
component at $E=6.43 \pm 0.04 \keV$ (rest frame) with $EW= 41 \pm 6 \eV$
corrected this issue.

\subsection{Joint Analysis of the {\em Suzaku} and {\em NuSTAR} Spectra}
\label{sec:broadband}

Including the {\it NuSTAR} data with our {\it Suzaku} data, along with their appropriate
cross-normalization factors, yields the highest S/N ever achieved
across the $0.7-79 \keV$ energy range.  This enables us to
probe the change in shape of the continuum at high energies, as well
as the contribution from reflection above $10 \keV$ and in the Fe K
band simultaneously.  

We plot the ratio of the {\it Suzaku}/XIS-FI, XIS-BI and HXD/PIN spectra, along with the
{\it NuSTAR}/FPMA and FPMB spectra to a power-law continuum modified by
Galactic photoabsorption in Fig.~\ref{fig:phpo_rat}.  The same
residuals detailed in \S\ref{sec:suzaku} are seen, but the curvature
above $10 \keV$ is now well-defined as a result of the higher S/N {\it
  NuSTAR} data.
 
\begin{figure}
\hbox{
\hspace{-1.0cm}
\includegraphics[width=0.4\textwidth,angle=270]{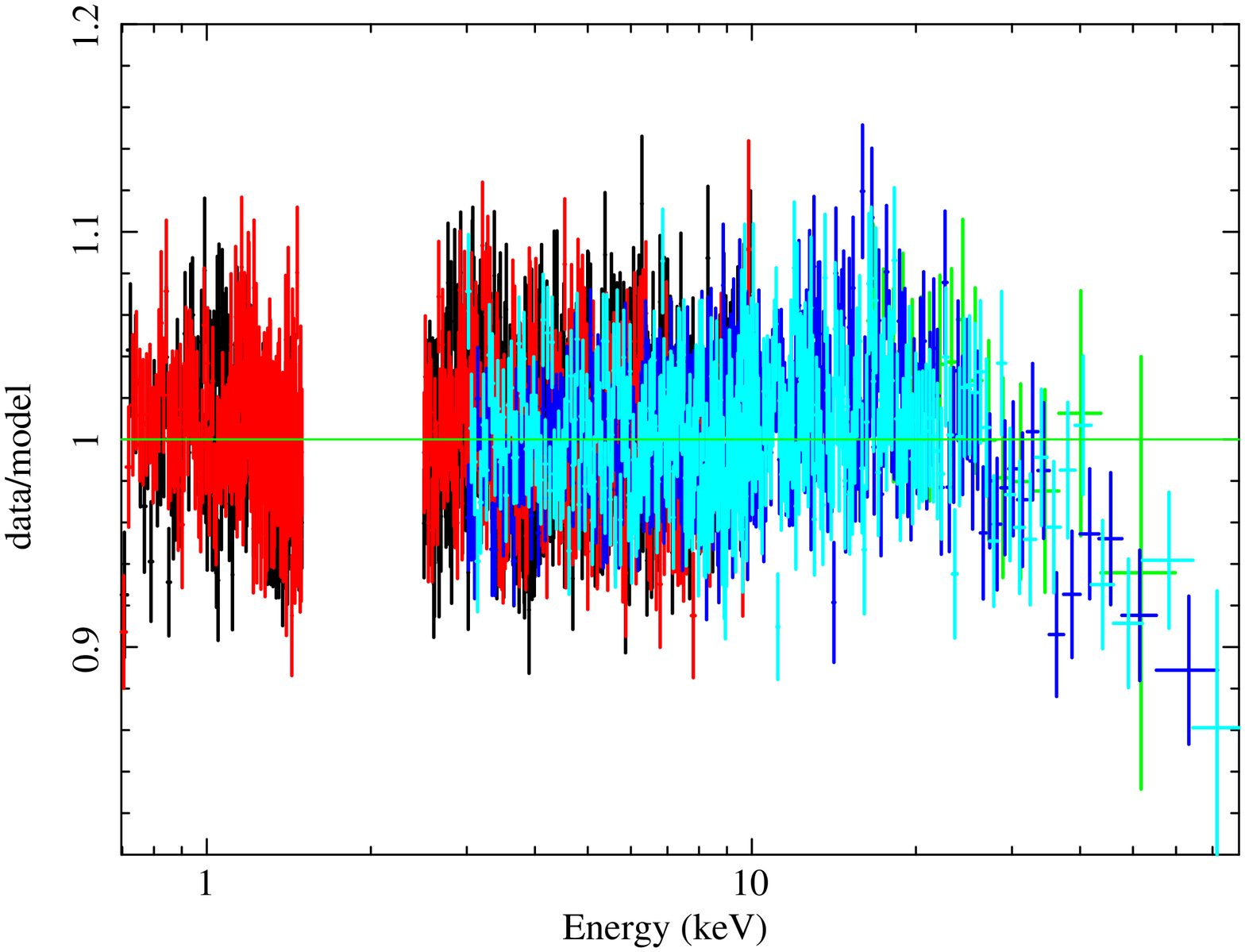}
\includegraphics[width=0.4\textwidth,angle=270]{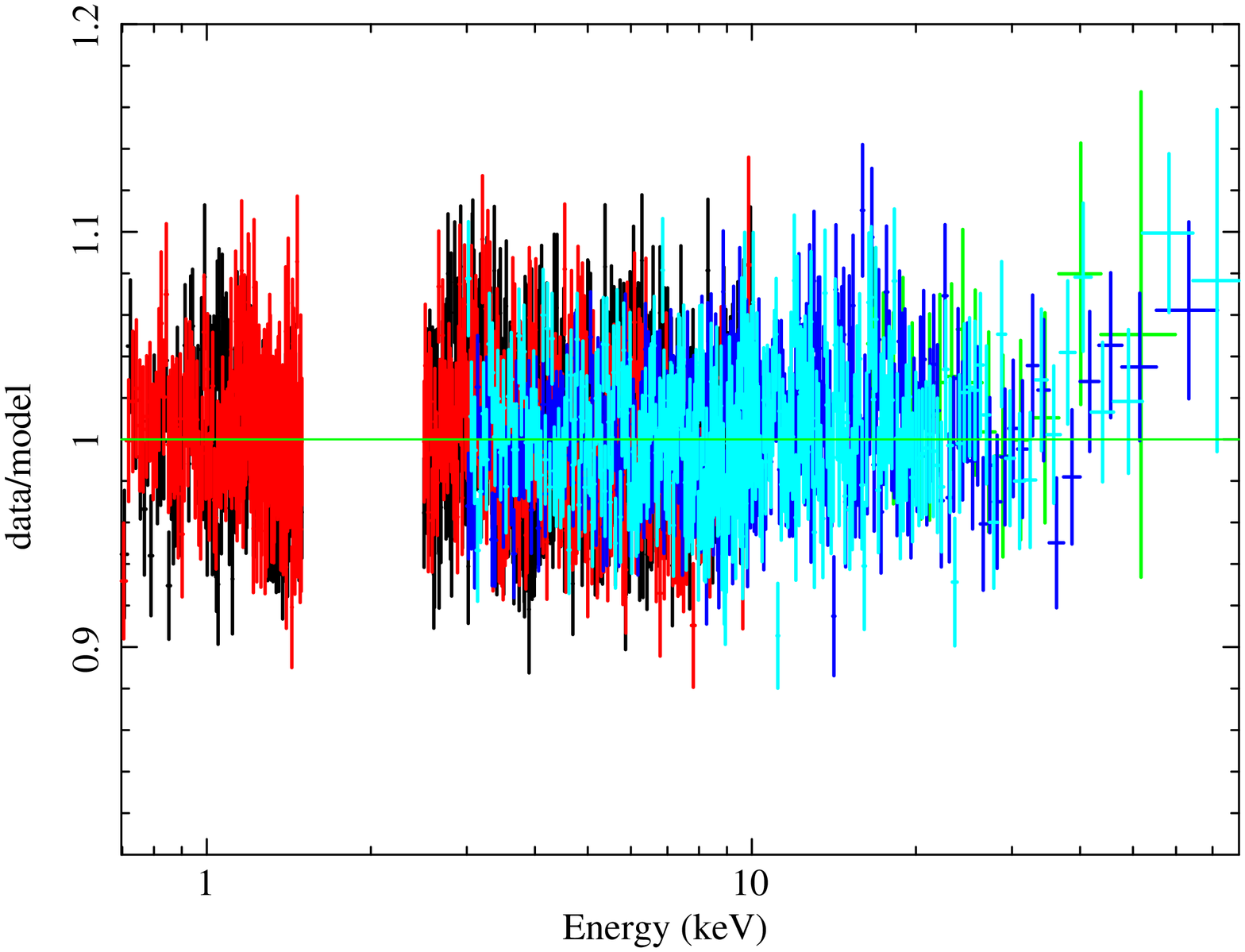}
}
\caption{{\small {\it Left:} Data-to-model ratio for the best-fitting {\tt
      xillver}+power-law model applied to the {\it Suzaku} data, now with data
    from {\it NuSTAR}/FPMA and FPMB added in and refitting performed.  Note the
    rollover of the spectra above $25 \keV$.  {\it Right:} The data-to-model ratio for Model~1,
    incorporating distant reflection through {\tt xillver}, a power-law
    continuum, and a high-energy cutoff.}}
\label{fig:rollover}
\end{figure}

A rollover at high energies is visually evident even from the {\it
  Suzaku}/PIN data, and becomes clearly evident when the FPMA and FPMB spectra
are added to the already-modeled {\it Suzaku} data, even though reflection is
already included in the model (see Fig.~\ref{fig:rollover}).  The presence of
this feature contributes to a global fit of
$\chi^2/\nu=4793/4341\,(1.10)$.  We now allow the
high-energy cutoff to the power-law and {\tt xillver} reflection
components to vary freely in
order to assess the robustness of such a feature over and above the distant
reflection, and, if present, to constrain its cutoff energy.  This approach
improves the fit to $\chi^2/\nu=4557/4340\,(1.05)$ and
greatly improves the data-to-model ratio at high
energies as well, visually.  The cutoff energy of the power-law is constrained
to $E_{\rm cut}=186 \pm 14 \keV$.  In comparison, paper I yielded
$E_{\rm cut}=178^{+74}_{-40} \keV$ with {\it NuSTAR} alone.  The residuals around the narrow Fe
K$\alpha$ line necessitate the addition of a broader Gaussian component,
as described in \S\ref{sec:suzaku}.  The best-fitting rest-frame energy of this line
is $E=6.46 \pm 0.08 \keV$, with equivalent width 
$EW=34 \pm 7 \eV$.  We will refer to this model as Model~1
throughout the rest of the paper.  

Given the high S/N out to $79 \keV$ achievable with {\it NuSTAR}, we can
consider more physically motivated models to describe the continuum emission than the more
phenomenological cutoff power-law.  To this end, in
paper I we employed the {\tt compTT} model of
\markcite{Titarchuk1994}{Titarchuk} (1994), parametrizing the continuum as being
produced by inverse Compton scattering of thermal disk photons off of
relativistic electrons situated above the disk in either a slab or spherical
geometry.  

\begin{figure}
\hbox{
\includegraphics[width=0.6\textwidth,angle=270]{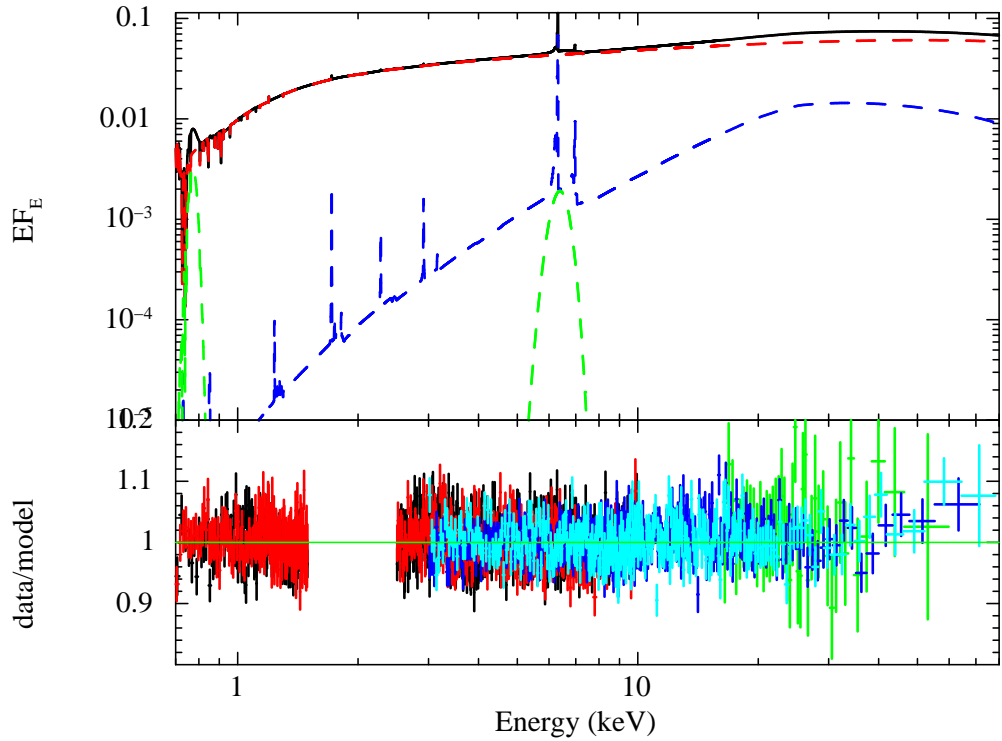}
}
\caption{{\small Best-fit components for Model~1.  The cutoff power-law
    continuum (modified by the warm absorber below $\sim2 \keV$) is shown in
    red, the distant reflection (also modified by the absorber) in
    dark blue, and the Gaussian lines of O\,{\sc
      vii} and Fe K are shown in green.  The summed model is in
    black.  Models~2-3 are virtually identical to Model~1 ---
    replacing the power-law with the {\tt compTT} Comptonization
    component --- and so are not included here.}}
\label{fig:mo1_eemodel}
\end{figure}

We adopt a similar approach in modeling the combined dataset here, applying a {\tt
  compTT} model component 
to the combined {\it Suzaku+NuSTAR} spectra, rather than a
phenomenological power-law continuum.  The rest of the model
components remained the same: the {\tt xillver} component used for the
reflected emission had its incident power-law spectral index frozen to
the best-fit value obtained in Model~1, much like the approach taken
using the {\tt compTT} model in paper I.
The spherical corona geometry is hereafter referred to as Model~2, while the
slab geometry is Model~3.  Plots of the contributions of individual
model components to the overall fit are shown in
Fig.~\ref{fig:mo1_eemodel}.  

Model~2 yields a goodness-of-fit
of $\chi^2/\nu=4550/4344\,(1.05)$, while Model~3 produces
$\chi^2/\nu=4555/4344\,(1.05)$; in comparison Model~1 comes to
$\chi^2/\nu=4558/4344\,(1.05)$.  None of the common parameter values change
significantly between models.  
The power-law of Model~1 finds its best fit with
$\Gamma=1.73 \pm 0.01$ and $E_{\rm cut}=186 \pm 14
\keV$, while the Comptonization components of Models~2-3 return
coronal temperatures and optical depths of $kT=50^{+6}_{-3} \keV$ and
$61 \pm 1 \keV$ and
$\tau=2.34^{+0.16}_{-0.21}$ and $0.68 \pm 0.02$, respectively.
In comparison, in paper I the {\it NuSTAR} data alone returned
$kT=33 \pm 6 \keV$ and $37^{+3}_{-6} \keV$ and
$\tau=3.41^{+0.58}_{-0.38}$ and $1.25^{+0.20}_{-0.10}$ for each
model.  We will discuss these differences further in \S\ref{sec:disc}

The Markov Chain Monte Carlo (MCMC) analysis used to measure the formal distribution of each
parameter across its available parameter space for each model was
performed using the Metropolis-Hastings algorithm (e.g.,
\markcite{Kashyap1998}{Kashyap} \& {Drake} 1998 and references therein) with four chains of
55,000 elements, in which the first
5000 elements from each were discarded as part of the ``burn-in''
period.  Each chain was started at a random seed in the global
parameter space using a diagonal covariance matrix with Gaussian
errors derived from the squares of the $1\sigma$ errors from the model fitting in
{\sc xspec}.  The rescaling factor of the covariance matrix was derived
using trial and error based on an initial estimate of $1/N^2$, where
$N$ is the number of free parameters in the fit.  The appropriate
rescaling factor was determined by ensuring that the fraction of
repeated values in a given chain was approximately $75\%$.  The
Gelman--Rubin convergence criterion of $\sim1$ was achieved for each
chain.\footnote{As outlined in the {\sc xspec} manual:
  http://heasarc.gsfc.nasa.gov/docs/xanadu/xspec/manual/XSchain.html}  

\begin{figure}
\hbox{
\centering
\includegraphics[width=0.35\textwidth,angle=270]{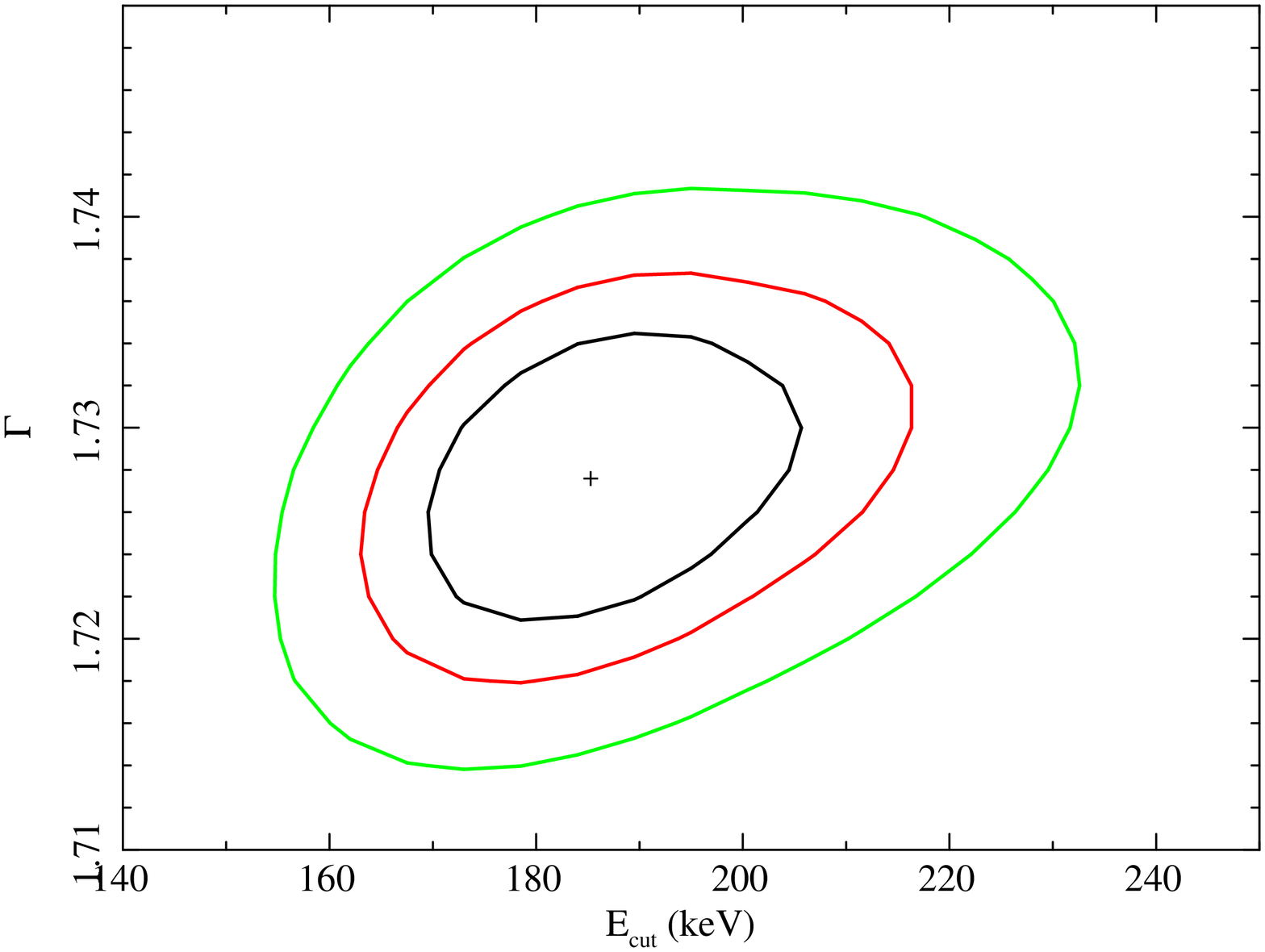}
\includegraphics[width=0.35\textwidth,angle=270]{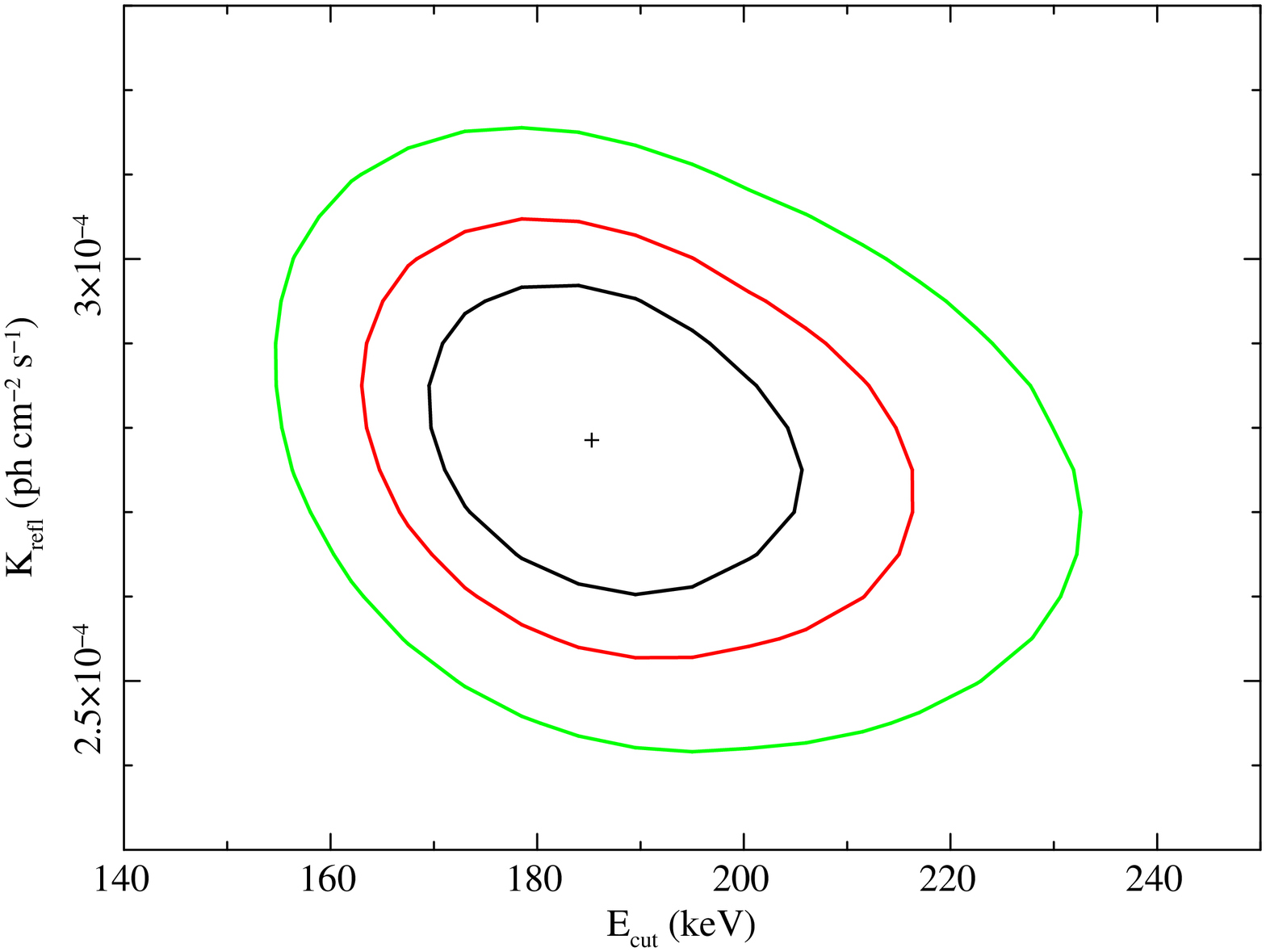}
}
\hbox{
\centering
\includegraphics[width=0.35\textwidth,angle=270]{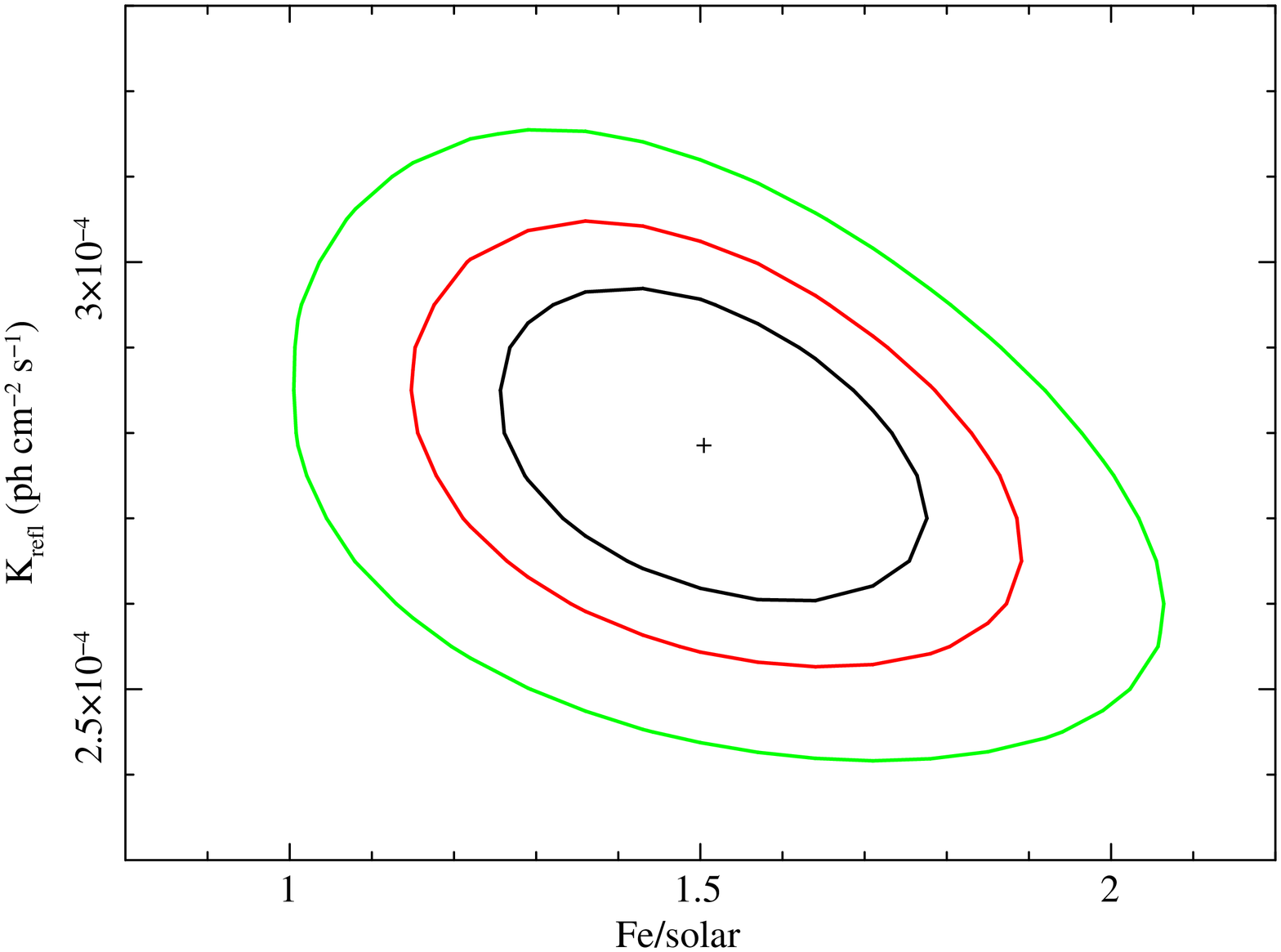}
\includegraphics[width=0.35\textwidth,angle=270]{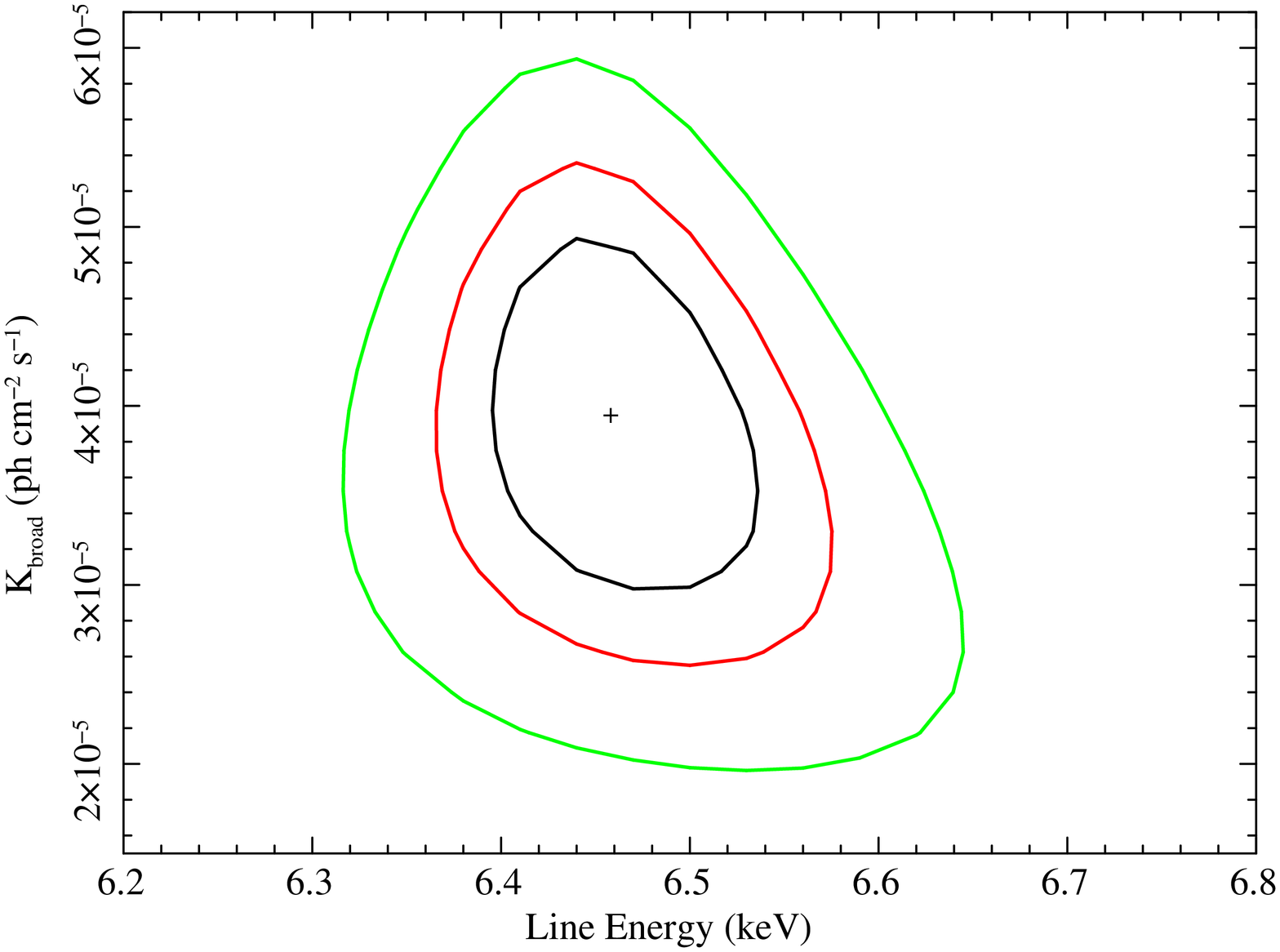}
}
\caption{\small{Contour plots from the MCMC analysis of Model~1, which
    show that, in each plot, the parameters are independently
  constrained and non-degenerate. 
    {\it Top left:} $67\%$, $90\%$ and $99\%$ confidence contours showing the constraints on the
photon index vs. cutoff energy of the power-law component.  {\it Top
  right:} The normalization of the distant reflector (a proxy for
reflection fraction) vs. the cutoff energy of the power-law
component. Normalization is in units of $\phpcmsqps$. 
  {\it Bottom left:} The iron abundance vs. normalization of the distant
  reflector.  
  {\it Bottom right:} The rest-frame line centroid energy
  vs. normalization for the broad Fe K$\alpha$ line component, also in
  $\phpcmsqps$.}}
\label{fig:mo1_contours}
\end{figure}

\begin{figure}
\hbox{
\centering
\includegraphics[width=0.35\textwidth,angle=270]{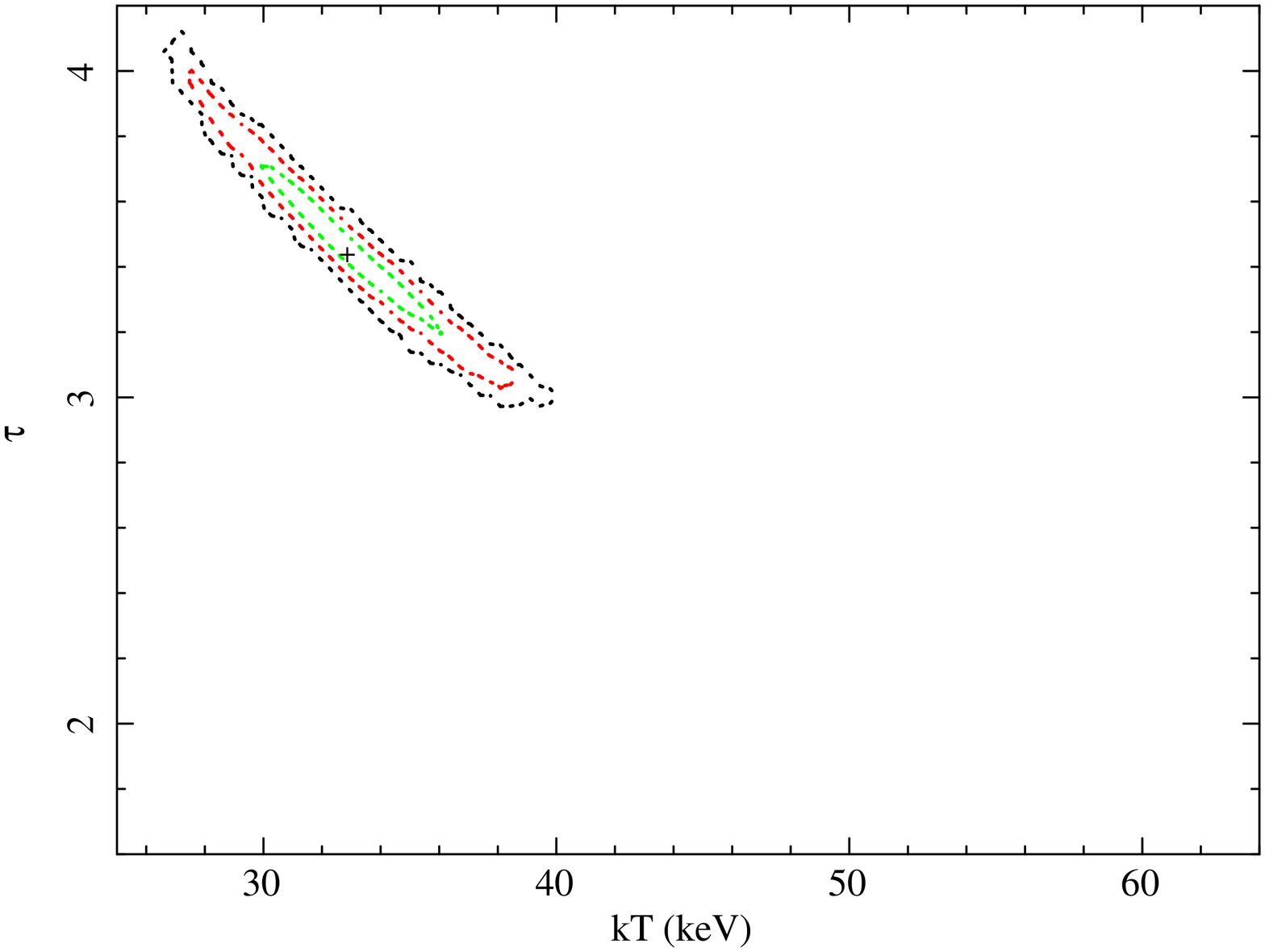}
\hspace{-8.3cm}
\includegraphics[width=0.35\textwidth,angle=270]{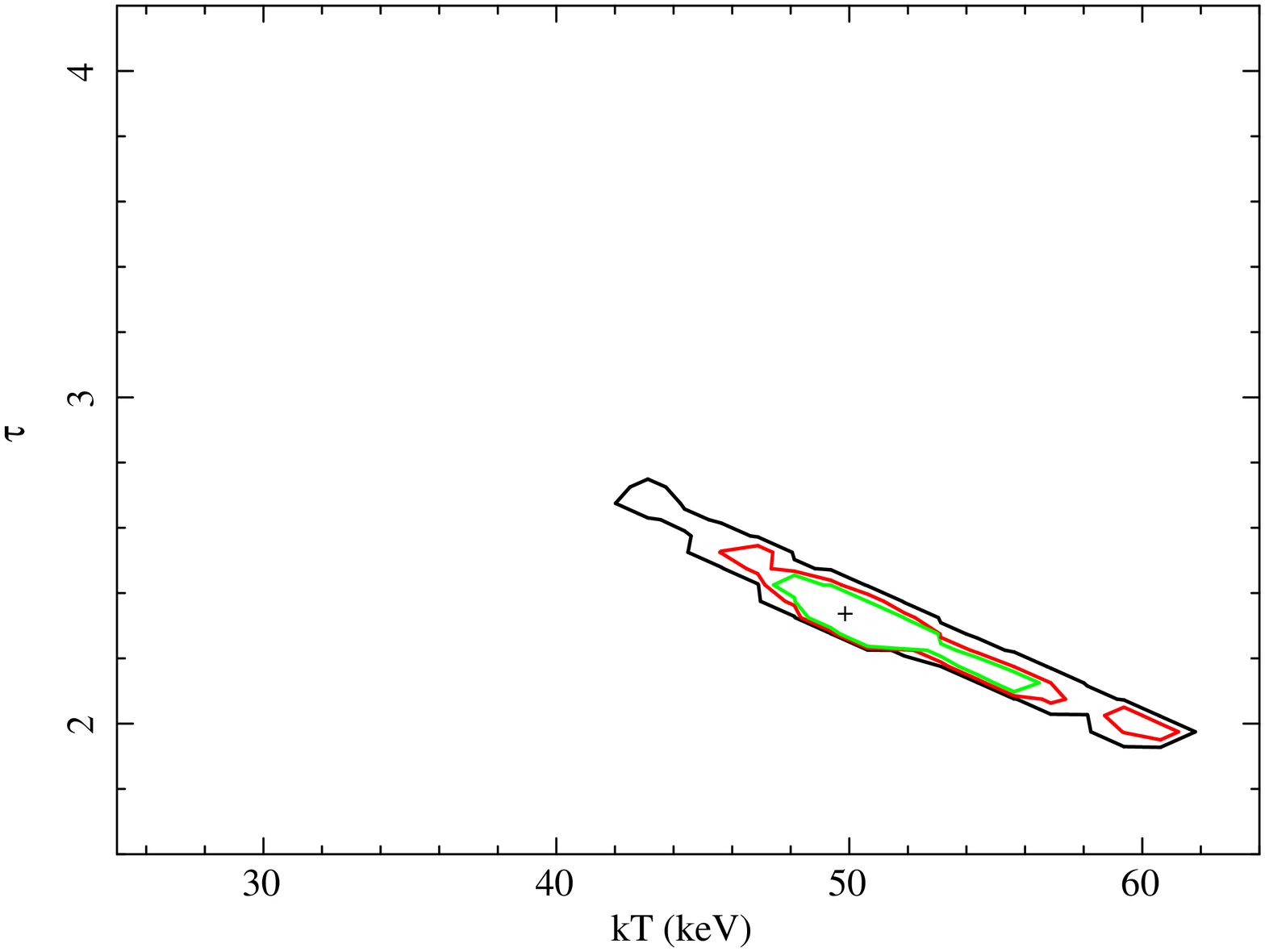}
\includegraphics[width=0.35\textwidth,angle=270]{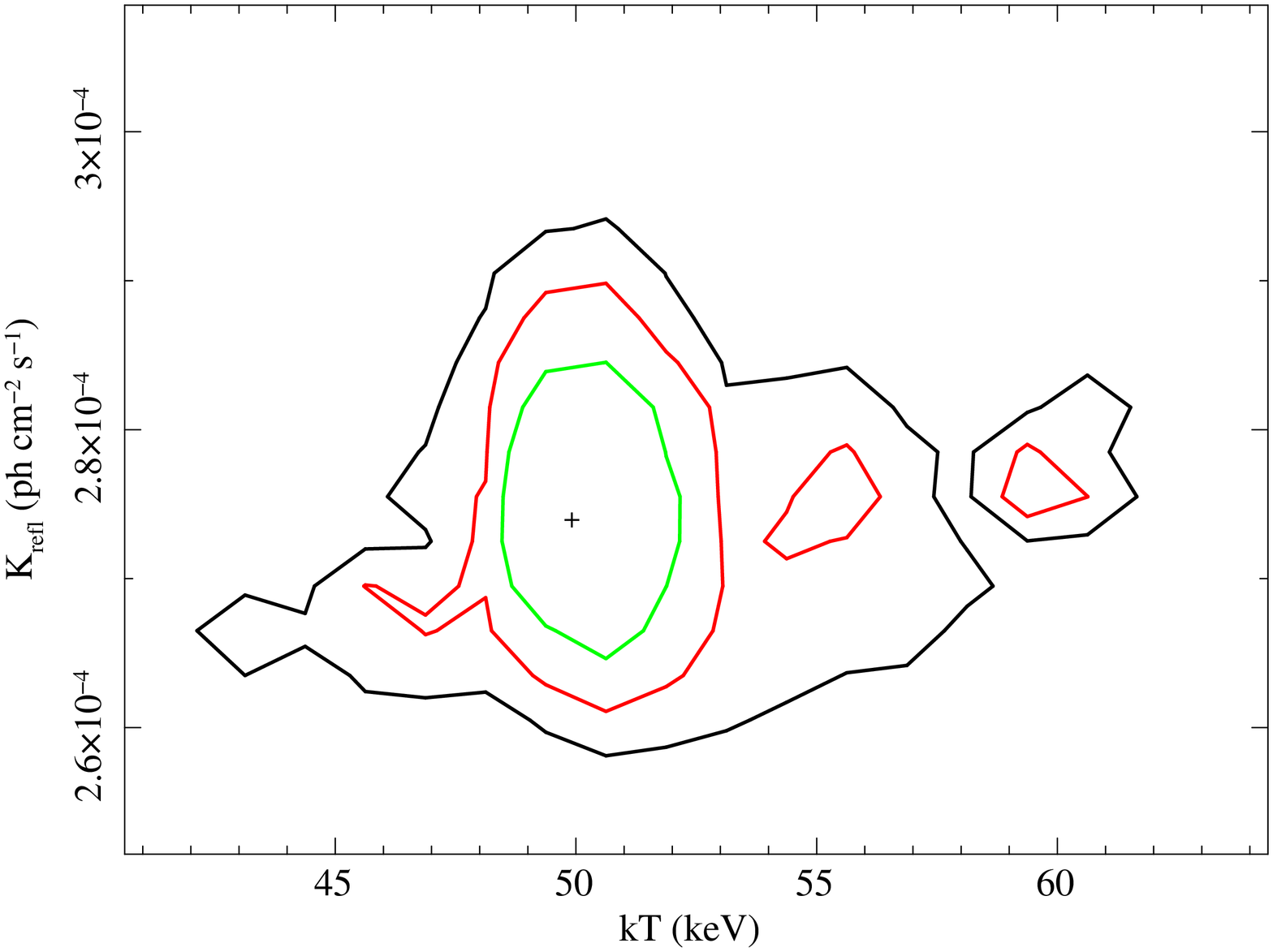}
}
\hbox{
\centering
\includegraphics[width=0.35\textwidth,angle=270]{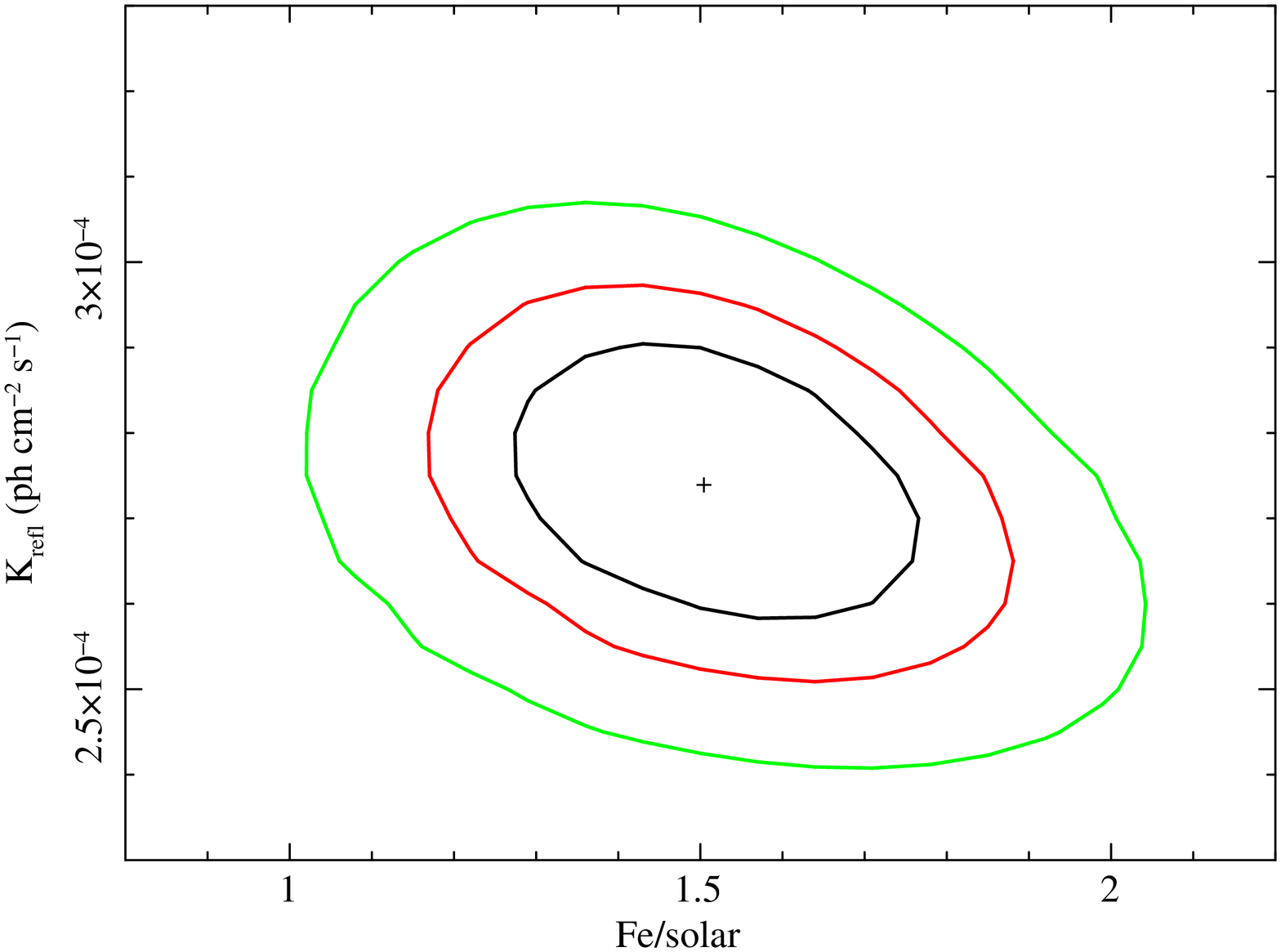}
\includegraphics[width=0.35\textwidth,angle=270]{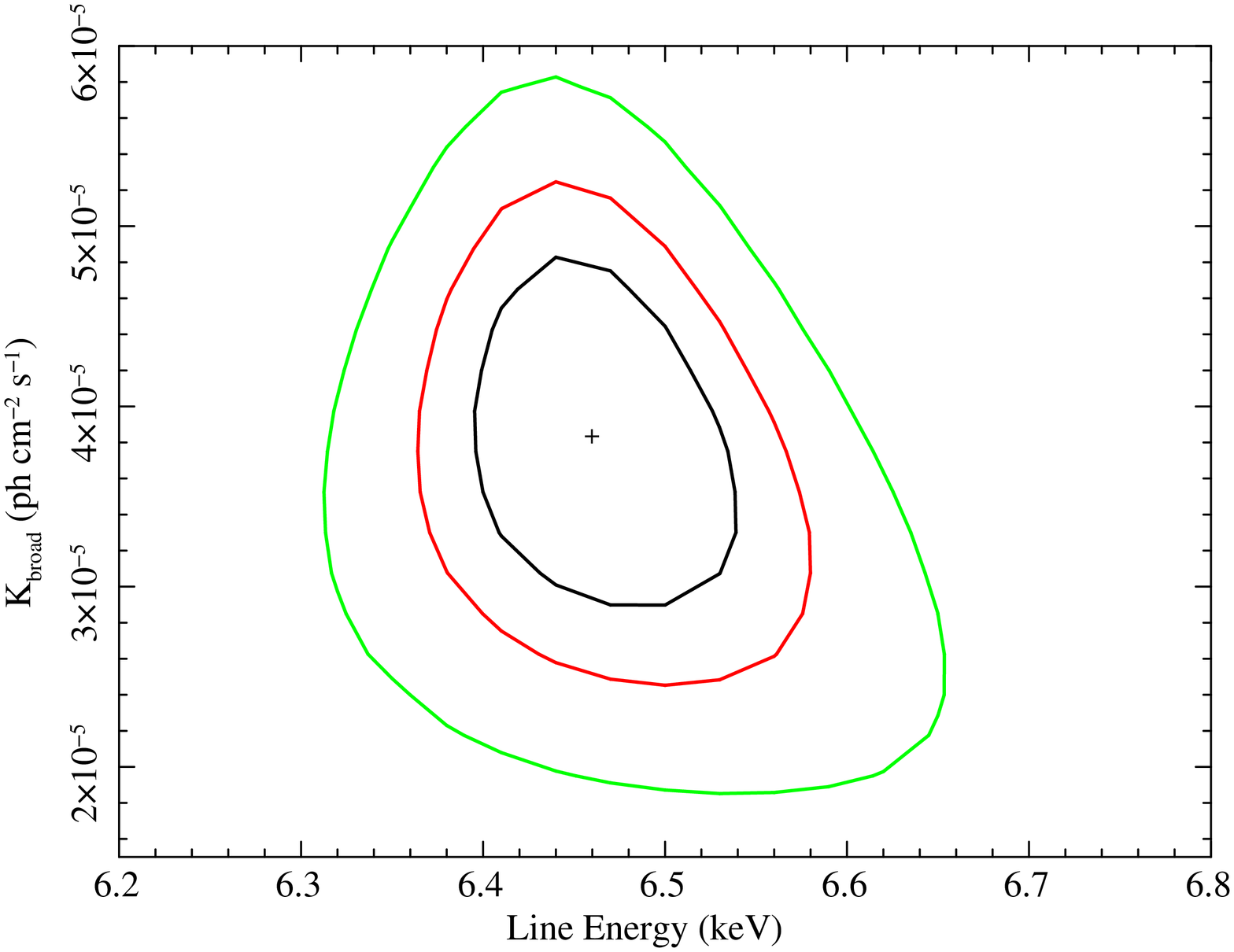}
}
\caption{{\small Contour plots from the MCMC analysis of Model~2.  
    {\it Top left:} $67\%$, $90\%$ and $99\%$ confidence contours showing the constraints on the
    coronal plasma temperature and optical depth for the spherical
    geometry for the {\it NuSTAR}-only (dotted) and {\it
      NuSTAR+Suzaku} (solid) data.  Though both parameters are
  constrained, there is still a clear degeneracy between the two.
  {\it Top right:} Normalization of the distant reflector vs. plasma
  temperature, with normalization in units of $\phpcmsqps$.  In this
  case the parameters are independently
  constrained and non-degenerate. 
  {\it Bottom left:} The iron abundance vs. normalization of the distant
  reflector.  
  {\it Bottom right:} The rest-frame line centroid energy
  vs. normalization for the broad Fe K$\alpha$ line component.}}
\label{fig:mo2_contours}
\end{figure}

\begin{figure}
\hbox{
\centering
\includegraphics[width=0.35\textwidth,angle=270]{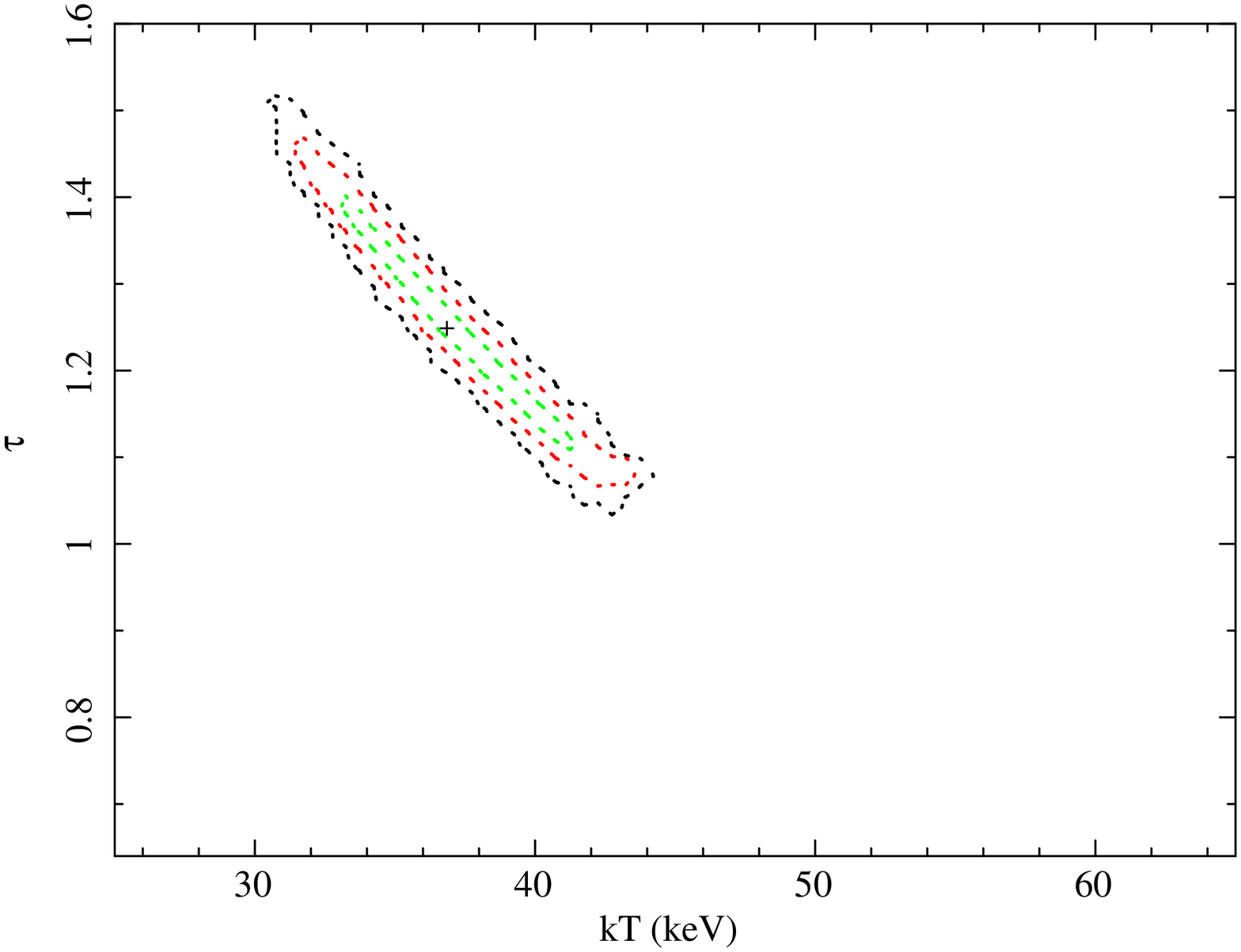}
\hspace{-7.95cm}
\includegraphics[width=0.35\textwidth,angle=270]{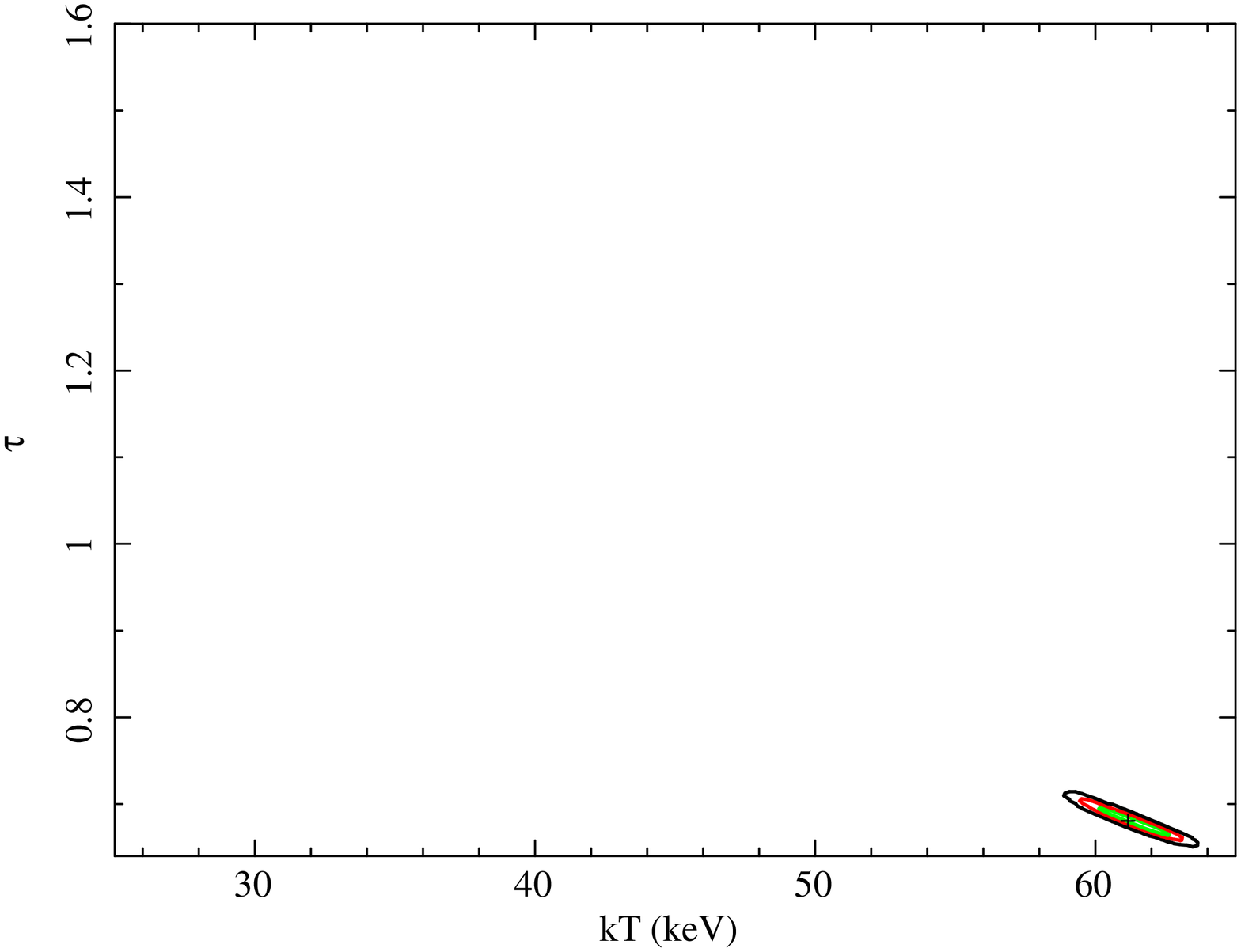}
\includegraphics[width=0.35\textwidth,angle=270]{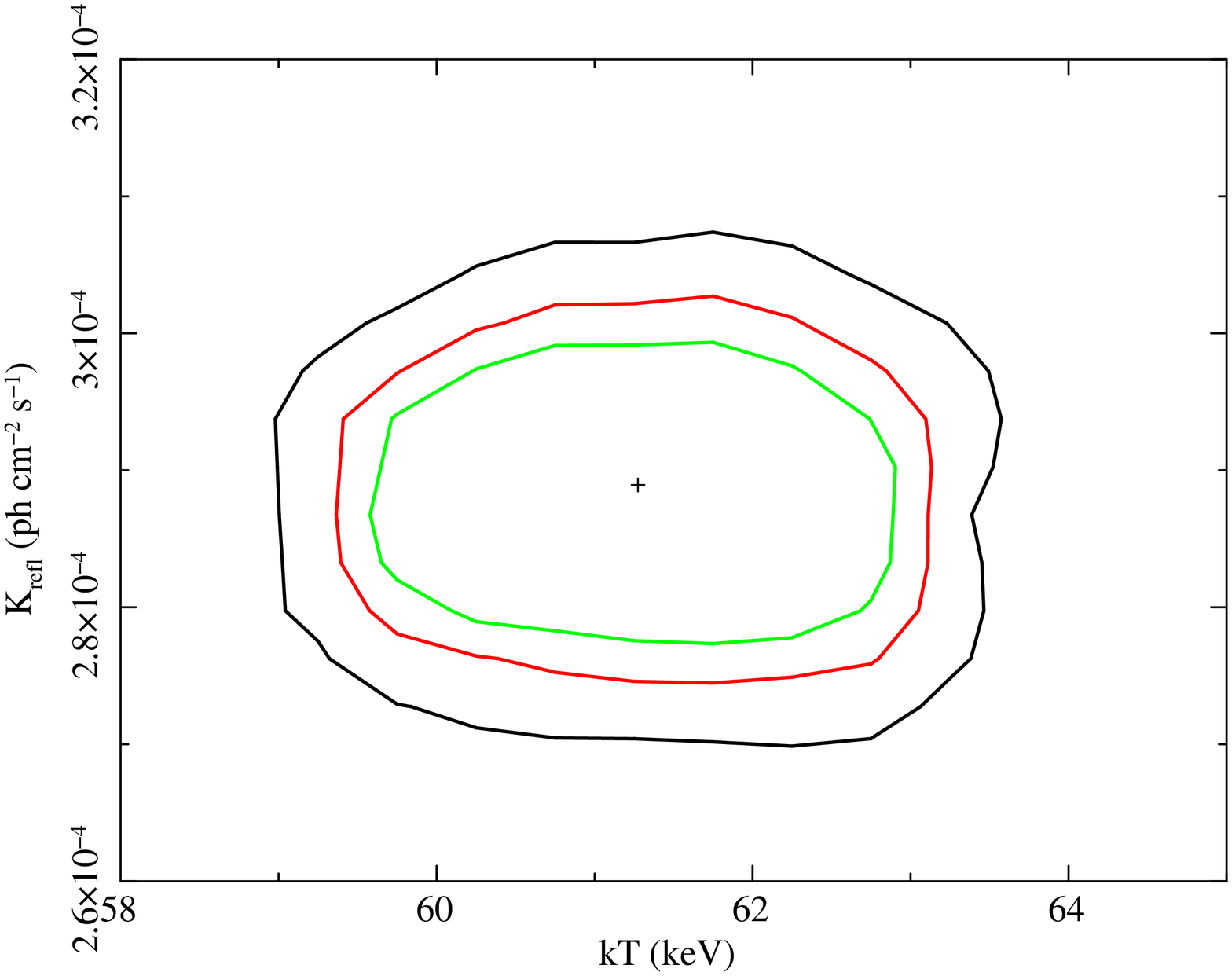}
}
\hbox{
\centering
\includegraphics[width=0.35\textwidth,angle=270]{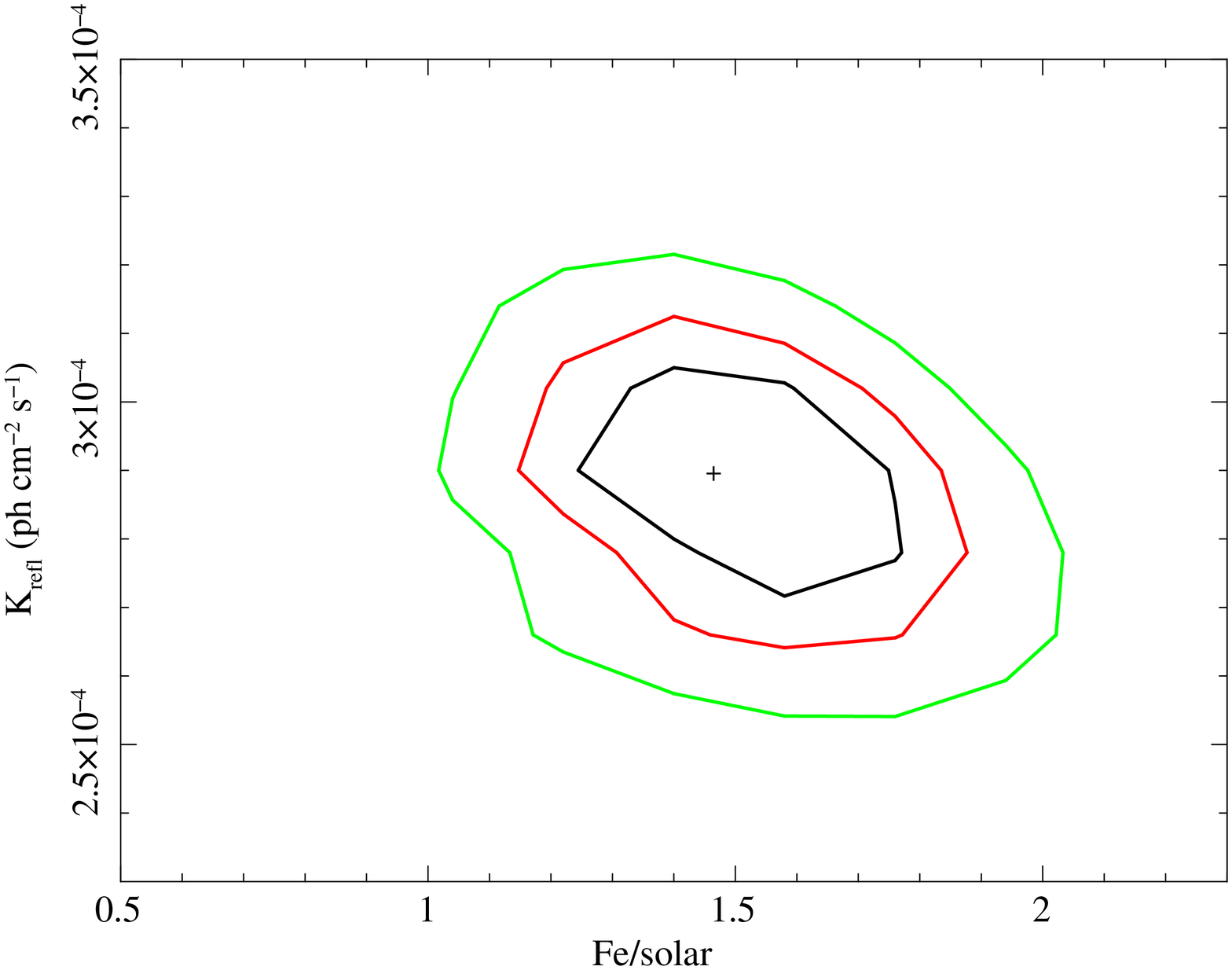}
\includegraphics[width=0.35\textwidth,angle=270]{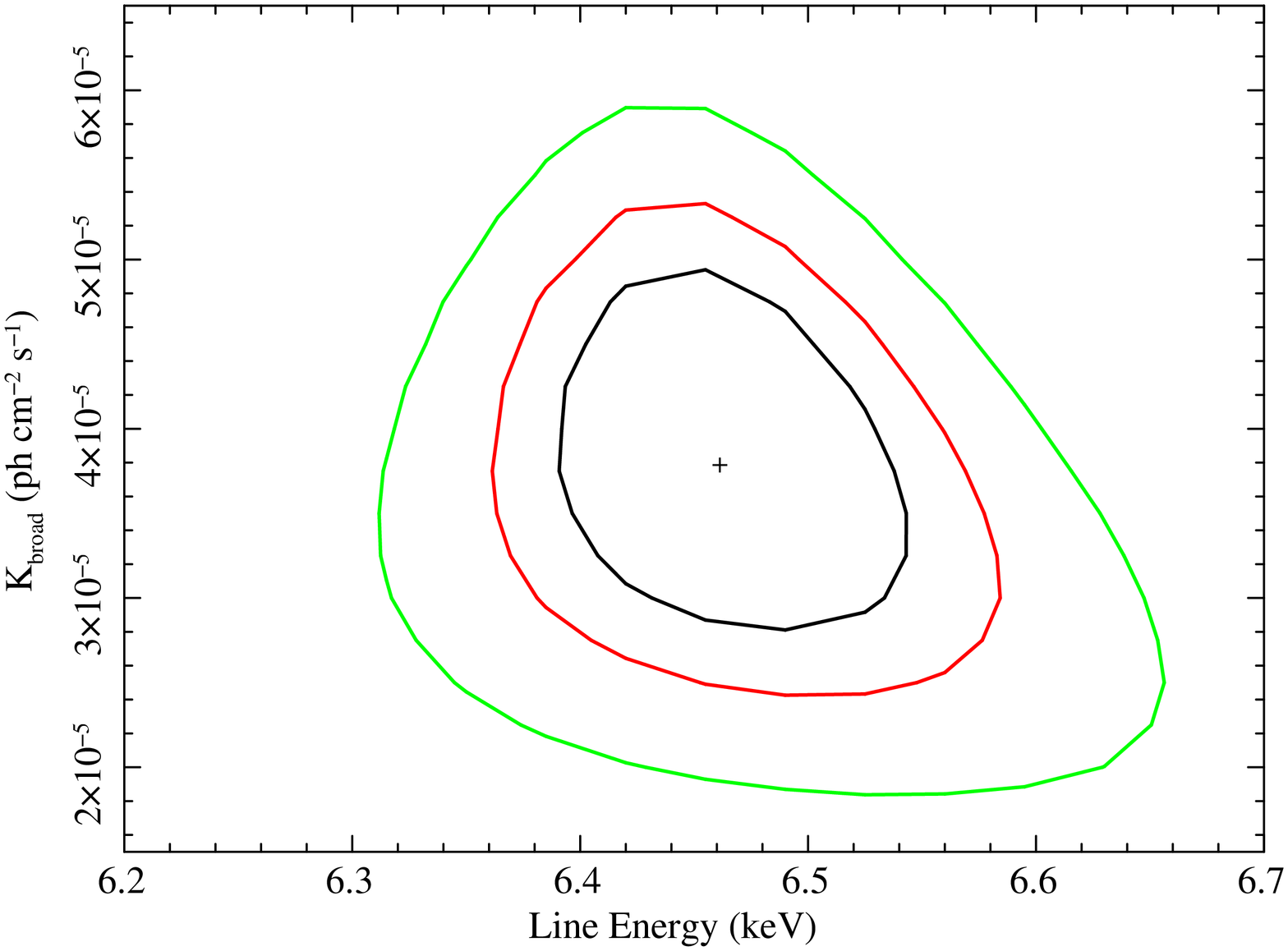}
}
\caption{{\small Contour plots from the MCMC analysis of Model~3.  
    {\it Top left:} $67\%$, $90\%$ and $99\%$ confidence contours showing the constraints on the
    coronal plasma temperature and optical depth for the spherical
    geometry for the {\it NuSTAR}-only (dotted) and {\it
      NuSTAR+Suzaku} (solid) data.  Though both parameters are
  constrained, there is still a clear degeneracy between the two.
  {\it Top right:} Normalization of the distant reflector vs. plasma
  temperature, with normalization in units of $\phpcmsqps$.  In this
  case the parameters are independently
  constrained and non-degenerate.
  {\it Bottom left:} The iron abundance vs. normalization of the distant
  reflector.  
  {\it Bottom right:} The rest-frame line centroid energy
  vs. normalization for the broad Fe K$\alpha$ line component.}}
\label{fig:mo3_contours}
\end{figure}

Once each chain run was completed, the four chains were loaded
back into {\sc xspec} to create a composite 200,000-element chain used
to extract $90\%$ confidence errors (shown in Table~\ref{tab:bigtab})
and to examine the distribution of each individual parameter in the
fit.  
We then generated probability density contours for the most
interesting pairs of parameters for each model, which are shown in
Figs.~\ref{fig:mo1_contours}, \ref{fig:mo2_contours} and
\ref{fig:mo3_contours}.  

The best-fit parameters and their MCMC-derived
$90\%$ confidence errors from all three models are shown in
Table~\ref{tab:bigtab}.  The total absorbed $2-10
\keV$ flux and luminosity of the model are $F_{2-10}=1.04 \times 10^{-10} \ergpcmsqps$
and $L_{2-10}=6.03 \times 10^{43} \ergps$, respectively, while their unabsorbed
values are $F_{2-10}=1.08 \times 10^{-10} \ergpcmsqps$ and $L_{2-10}=6.26 \times
10^{43} \ergps$.

{\small
\begin{table}
\begin{tabular}{|lllll|}\hline\hline
{\bf Component} & {\bf Parameter (units)} & {\bf Model~1} & {\bf Model~2} & {\bf
  Model~3} \\
\hline \hline
{\tt TBabs} & $N_{\rm H}\,(\pcmsq)$ & $4.61 \times 10^{20}(f)$ & $4.61 \times
10^{20}(f)$ & $4.61 \times 10^{20}(f)$ \\
{\sc xstar} grid & $N_{\rm H}\,(\pcmsq)$ & $6.03 \pm 0.13 \times
10^{21}$ & $6.02 \pm 0.13 \times 10^{21}$ & $6.00^{+0.12}_{-0.13} \times 10^{21}$ \\
   & ${\rm log}\, \xi\,(\ergcmps)$ & $0.73 \pm 0.02$ & $0.73 \pm 0.02$
   & $0.73 \pm 0.02$ \\
\hline
{\tt zpo} & $\Gamma$ & $1.73 \pm 0.01$ & $1.73(f)$ & $1.73(f)$ \\
          & $E_{\rm cut}\,(\keV)$ & $186^{+14}_{-14}$ & $---$ & $---$ \\
          & $K_{\rm po}\,(\phpcmsqps)$ & $2.82 \pm 0.03 \times
          10^{-2}$ & $---$ & $---$ \\
\hline
{\tt compTT} & $kT_e\,(\keV)$ & $---$ & $50^{+6}_{-3}$ & $61 \pm 1$ \\
             & $\tau$ & $---$ & $2.34^{+0.16}_{-0.21}$ & $0.68 \pm 0.02$ \\
             & $K_{\rm comptt}\,(\phpcmsqps)$ & $---$ &
             $5.46^{+0.38}_{-0.54} \times 10^{-3}$ & $4.46 \pm 0.11 \times 10^{-3}$ \\
\hline
{\tt xillver} & Fe/solar & $1.51^{+0.29}_{-0.28}$ & $1.51^{+0.28}_{-0.27}$ & $1.50^{+0.28}_{-0.27}$ \\
          & $\Gamma$ & $1.73*$ & $1.73(f)$ & $1.73(f)$ \\
          & $K_{\rm refl}\,(\phpcmsqps)$ & $2.79 \pm 0.20
          \times 10^{-4}$ & $2.74 \pm 0.18 \times 10^{-4}$ &
          $2.89 \pm 0.18 \times 10^{-4}$ \\
\hline
{\tt zgauss} & $E_1\,(\keV)$ & $0.78 \pm 0.01$ & $0.78 \pm 0.01$ &$0.78 \pm 0.01$ \\
             & $\sigma_1\,(\keV)$ & $2.04^{+0.50}_{-0.67} \times 10^{-2}$ & $1.97^{+0.49}_{-0.57} \times 10^{-2}$
  & $2.17^{+0.43}_{-0.70} \times 10^{-2}$ \\
             & $K_1\,(\phpcmsqps)$ & $1.52^{+0.19}_{-0.25} \times
             10^{-3}$ & $1.48^{+0.23}_{-0.22} \times 10^{-3}$ & $1.57^{+0.21}_{-0.25} \times 10^{-3}$ \\
             & $EW_1\,(\eV)$ & $36^{+5}_{-6}$ & $37 \pm 6$ & $37^{+5}_{-6}$ \\
{\tt zgauss} & $E_2\,(\keV)$ & $6.46^{+0.08}_{-0.07}$ & $6.46^{+0.09}_{-0.07}$ & $6.46^{+0.08}_{-0.07}$ \\
             & $\sigma_2\,(\keV)$ & $0.33^{+0.08}_{-0.07}$ & $0.33^{+0.09}_{-0.07}$ & $0.33^{+0.09}_{-0.07}$ \\
             & $K_2\,(\phpcmsqps)$ & $3.95^{+1.05}_{-1.08} \times 10^{-5}$ & $3.84^{+1.05}_{-1.03} \times
  10^{-5}$ & $3.91 \pm 1.04 \times 10^{-5}$ \\
             & $EW_2\,(\eV)$ & $34^{+8}_{-7}$ & $27^{+7}_{-6}$ & $26
             \pm 7$ \\
\hline \hline
Final fit & & $4558/4344\,(1.05)$ & $4550/4344\,(1.05)$ & $4555/4344\,(1.05)$ \\
\hline \hline
\end{tabular}
\hspace{-1.0cm}
\caption{\small{Best-fit parameters, their values and uncertainties (to $90\%$
    confidence) for Models~1-3 fit to the {\it Suzaku+NuSTAR} data
    (see text for model details).  Parameters marked
    with an ({\it f}) are held fixed in the fit, while those marked with an (*) are
    tied to another parameter (see text).  Redshifts are fixed at the
    cosmological value for IC~4329A ($z=0.0161$) unless otherwise specified.
    The warm absorber is assumed to totally cover the source and to have solar
    abundances.  The {\tt xillver} model for distant reflection has a photon
    index tied to that of the primary continuum and an ionization fixed to its
    lowest possible value (log $\xi=0$).  Its inclination angle to the line of
    sight is fixed to $60 \degmark$.}}
\label{tab:bigtab}
\end{table}
}

\subsection{Spectral Variability}
\label{sec:spec_var}

Our simultaneous {\it Suzaku} and {\it NuSTAR} observations of
IC~4329A do not show marked variability on short timescales, as 
exhibited by
many other AGN (e.g., NGC~1365, \markcite{Risaliti2013}{Risaliti}
{et~al.} 2013).  Nonetheless, it is useful to examine time-sliced
spectra from the highest and lowest flux states of the source during
our campaign.  This exercise yields insight into the changes in certain spectral
components driving the change in flux, i.e., to what degree these flux
changes are caused by variations in the continuum, absorption or
reflection components.

\begin{figure}
\hbox{
\includegraphics[width=0.6\textwidth,angle=270]{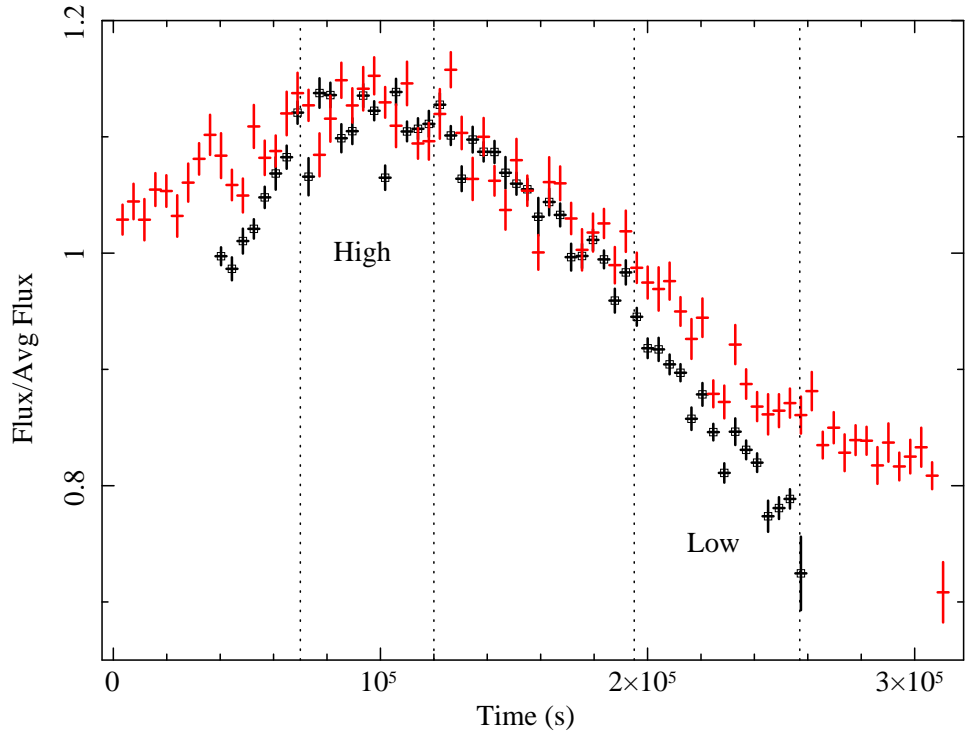}
}
\caption{{\small High- and low-flux time intervals common to the {\it
      Suzaku} and {\it NuSTAR} observations.  {\it Suzaku}/XIS~3
    data ($0.7-10 \keV$) are shown in black (with square markers), {\it NuSTAR}/FPMA data ($3-79
    \keV$) in red (crosses), each normalized by their
    mean count rate to show the similarity in shape between the light curves.}}
\label{fig:high_low_times}
\end{figure}

\begin{figure}
\hbox{
\includegraphics[width=0.6\textwidth,angle=270]{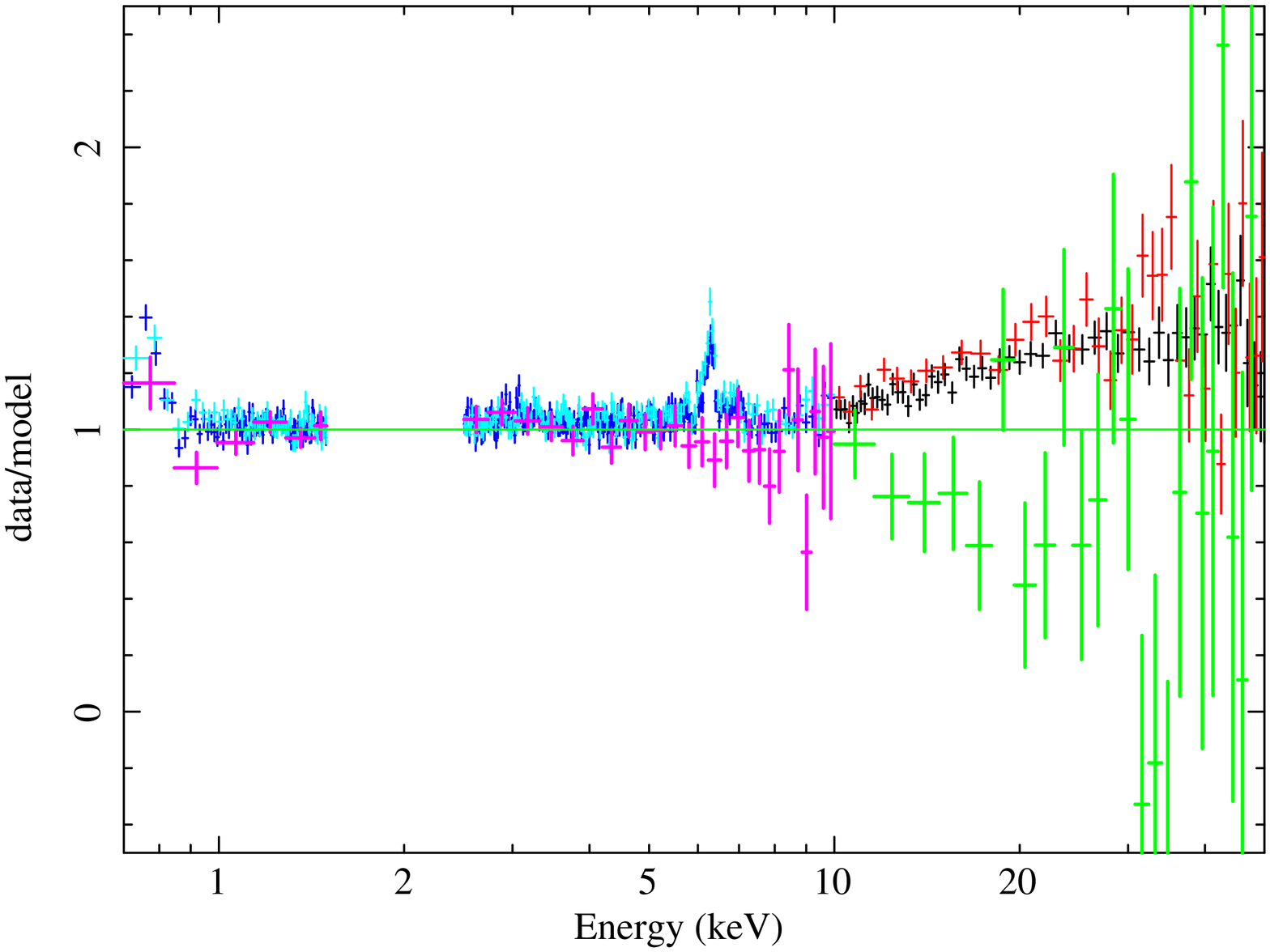}
}
\caption{{\small High-flux, low-flux and difference spectra from {\it Suzaku}/XIS-FI
    and {\it NuSTAR}/FPMA ratioed against a photoabsorbed power-law.  Note the
    lack of
    reflection features evident in the residuals for the difference spectra.
    This is a strong indication that the power-law is driving the
    spectral changes over the course of the
    observation.  The {\it Suzaku} high-flux spectrum is in dark blue, the low-flux
    spectrum is in light blue and the difference spectrum is in magenta.  The
    {\it NuSTAR} high-flux spectrum is in black, the low-flux spectrum is in red
and the difference spectrum is in green. The difference spectra are
more coarsely binned for visualization purposes, and have larger
uncertainties indicative of the marginal differences between the high
and low flux states of IC~4329A, especially above $10 \keV$.}}
\label{fig:high_low_diff_norefl}
\end{figure}

We extracted simultaneous high- and low-flux spectra from the time intervals in each
observation ({\it Suzaku}/XIS-FI and {\it NuSTAR}/FPMA; FPMB are not shown for
clarity, but are virtually identical to FPMA).  These spectra are shown in
Fig.~\ref{fig:high_low_times}.  The high-flux
state spectra totaled $46 \ks$ of exposure time and represent the highest source flux
with simultaneous data from the two telescopes.  The low-flux state totaled $63
\ks$ of exposure time, and represents the lowest source flux with simultaneous data
from both telescopes.  Difference spectra
were created by subtracting the low-flux spectra from the high-flux
spectra in both {\it Suzaku}/XIS-FI and {\it NuSTAR}/FPMA.  We show
all six spectra plotted against the best-fit power-law modified by Galactic
photoabsorption in Fig.~\ref{fig:high_low_diff_norefl}.  Note that the
difference spectra show no change from the high- and low-state spectra below
$\sim2 \keV$, indicating the constancy of the warm absorber during our
observations.  Close inspection of Fig.~\ref{fig:high_low_diff_norefl} also
reveals that the reflection features (Fe K band and Compton hump above $10
\keV$) are slightly more prominent during the low-flux state when the power-law
emission is minimized.  The difference spectra show no residual reflection
features, implying that these are constant in flux over the course of the
observation.  We can therefore infer that changes in only the power-law flux drive
the slight spectral evolution that takes place as IC~4329A transitions from a
higher-flux state to a lower-flux state during our campaign.

The source is softer when brighter, as is typical in many actively
accreting AGN:
$\Gamma=1.73 \pm 0.01$ in the high-flux state ($K_{\rm po}=3.01 \times 10^{-2}
\phpcmsqps$) vs. $\Gamma=1.68 \pm 0.01$ in the low-flux state ($K_{\rm po}=2.23 \times 10^{-2}
\phpcmsqps$).  This observed relation is thought to arise due to a
correlation between the mass accretion rate and power-law slope,
wherein higher accretion rates result in steeper spectra.  This result can
be explained if the Compton amplification factor decreases
proportionally with the accretion rate in AGN (e.g.,
\markcite{Sobolewska2009}{Sobolewska} \& {Papadakis} 2009 and references therein), though note all AGN
display this behavior (e.g., NGC~4151; \markcite{Lubinski2010}{Lubi{\'n}ski} {et~al.} 2010).

The cutoff energy of the power-law in our data does not vary
significantly from its best-fit value between the high- and low-flux
spectra, and cannot be constrained in the difference spectrum.  The
difference spectrum is well fit by a power-law modified by
both Galactic and intrinsic absorption, much like that described in the Model~1.
This power-law has a slope of $\Gamma=1.79 \pm 0.02$ and a normalization of
$K_{\rm po}=7.84 \times 10^{-3} \phpcmsqps$.

\section{Discussion}
\label{sec:disc}

\subsection{Summary}
\label{sec:summary}

In our deep observation of IC~4329A, performed simultaneously with {\it
  Suzaku} and {\it NuSTAR}, we are able to robustly separate the continuum, absorption and
distant reflection components in the spectrum using the broad energy range of
our observations.  The results of our analysis can be summarized as
follows:

\begin{itemize}
\item{IC~4329A was viewed in a flux state near its historical
average, and displayed little variability on short timescales and 
$\sim30\%$ variability over the course of the campaign, {\bf as has been
found in previous X-ray observations}.}  
\item{While we were able to place a strong constraint on the
presence of a broadened Fe K$\alpha$ line in the data ($EW=34^{+6}_{-9} \eV$ in
Model~1, the highest equivalent width seen in our modeling), we were not able to
constrain any of the parameters when a relativistic line model was applied to
the data.  As such, it is not possible to derive any constraints on the spin of
the black hole in IC~4329A using these observations.}
\item{We have made the most accurate, precise
measurement of the high-energy cutoff of the X-ray emission to date: $E_{\rm
  cut}=186 \pm 14 \keV$.  This measurement improves on that made
with {\it NuSTAR} alone in paper I ($E_{\rm cut}=178^{+74}_{-40}
\keV$), demonstrating the necessity of obtaining
high-S/N, broadband X-ray data in order to determine the properties of
the corona.}
\item{Using data from both {\it Suzaku} and {\it NuSTAR}, we
derive $kT=50^{+6}_{-3} \keV$ and $\tau=2.34^{+0.16}_{-0.21}$ for the
spherical geometry, with $kT=61 \pm 1 \keV$ and
$\tau=0.68 \pm 0.02$ for the slab geometry.}  
\end{itemize}

\subsection{Understanding the Corona}
\label{sec:corona}

It is important to establish the continuum level in AGN in
order to determine their overall energy budget, and the high-S/N,
broadband X-ray spectra we have obtained using {\it Suzaku} and {\it
  NuSTAR} simultaneously enable us to disentangle the continuum,
absorption and reflection signatures more accurately than we are able
to with either observatory alone. 

Our spectral and timing analyses of the simultaneous {\it Suzaku} and
{\it NuSTAR} observations of IC~4329A demonstrate that changes in
the continuum flux are responsible for the modest, secular changes in
the overall source flux that we detect.  The high-S/N, broadband
spectra enable us to determine that neither the absorption nor
reflection components show any significant variability over the
course of our observing campaign.  Though this result was hinted at in
paper I, the addition of the {\it Suzaku} data to the analysis
confirms the lack of absorption variability, in particular.  Referencing Model~1, we note
that the power-law normalization decreases by $\sim30\%$ between the
high- and low-flux states during our observations (\S\ref{sec:spec_var}); similarly, the
overall source flux decreases by approximately the same amount
(\S\ref{sec:timing}).  The variation is more pronounced at
energies below $\sim10 \keV$ (Fig.~\ref{fig:high_low_diff_norefl}), in
keeping with our examination of the RMS variability spectrum
(Fig.~\ref{fig:fvar}).

We note that the spectrum is rather hard: the cutoff power-law fit in
Model~1 returns a photon index of $\Gamma=1.73 \pm 0.01$, also
confirming the results from paper I.  This is
consistent with several recent measurements taken with {\it Chandra}
and {\it XMM} (average $\Gamma=1.73$;
\markcite{McKernan2004,Steenbrugge2005,Markowitz2006}{McKernan} \& {Yaqoob} 2004; {Steenbrugge} {et~al.} 2005; {Markowitz}, {Reeves}, \&  {Braito} 2006), but
inconsistent with the joint {\it
  XMM+INTEGRAL} spectral fitting performed by \markcite{Molina2009}{Molina} {et~al.} (2009)
($\Gamma=1.81 \pm 0.03$).  We
note, however, that the data reported by Molina \etal were not
obtained simultaneously.  
Spectral fitting returned a significantly softer index in previous
epochs as well; between 1995
and 2001, the average reported spectral slope was $\Gamma=1.91$ with a
range between $\Gamma=1.83-2.0$
\markcite{Madejski1995,Cappi1996,Perola1999,Done2000,Gondoin2001}({Madejski} {et~al.} 1995; {Cappi} {et~al.} 1996; {Perola} {et~al.} 1999; {Done} {et~al.} 2000; {Gondoin} {et~al.} 2001).
Caution should be used when measuring the power-law slope using only
data $\leq10 \keV$, as the true slope of the continuum is best
assessed over a much broader energy band extending out to higher
energies where the continuum is more dominant.  We also note that the
quality of our data at high energies with {\it NuSTAR} now far surpasses that
of the spectra obtained with {\it RXTE, BeppoSAX, CGRO} or {\it
  INTEGRAL}.  Even so, these differences in measured power-law slope
may indicate that IC~4329A undergoes significant
coronal variability on $\leq$years-long timescales.  

It would be interesting to investigate whether the cutoff energy of
the power-law shows similar variations to the spectral index, but
unfortunately the constraints placed on this parameter historically
are too loose to enable this test.  In all of the observations prior
to 2009, either the cutoff energy was fixed to the $E_{\rm
  cut}=270^{+167}_{-80} \keV$ result obtained by \markcite{Perola1999}{Perola} {et~al.} (1999),
or did not improve on this result (e.g., $E_{\rm cut} \geq 100 \keV$,
\markcite{Madejski1995}{Madejski} {et~al.} 1995; $E_{\rm cut}=270 \pm 120
\keV$, \markcite{Gondoin2001}{Gondoin} {et~al.} 2001).  Our result ($E_{\rm cut}=186 \pm 14
\keV$, consistent with yet more precise than the measurement from
paper I) is consistent with that
obtained by \markcite{Molina2009}{Molina} {et~al.} (2009), as discussed in
\S~\ref{sec:broadband}.  It is also at the median point of the
high-energy power-law cutoffs that have been measured in Seyfert AGN
thus far with {\it NuSTAR}.  Others include Ark~120 ($\geq190 \keV$,
\markcite{Matt2014}{Matt} {et~al.} 2014), SWIFT~J2127.4+5654 ($108^{+11}_{-10} \keV$,
\markcite{Marinucci2014}{Marinucci} {et~al.} 2014) and Mrk~335 ($\geq153 \keV$, Parker \etal,
  submitted). 

Owing to the high quality and broadband energy coverage of our data, we were able to reach
beyond the phenomenological power-law representation of the continuum and consider
more physical models, following our work with the {\it NuSTAR} data
alone in paper I.  Models~2-3 provide
roughly equivalent statistical fits to the data,
incorporating a {\tt compTT} model that parametrizes the temperature,
optical depth and geometry of the electron
plasma, as well as an {\tt xillver} model that assumes a neutral slab of
material inclined at $60 \degmark$ to our line of sight and leaves the
reflected flux from the disk and its iron abundance as free
parameters.  We also added in two Gaussian
components to represent (1) a blend of the resonance, intercombination and
forbidden O\,{\sc vii} emission lines, and (2) a
contribution from a broad Fe K$\alpha$ line from the inner disk.  

\begin{figure}[H]
\hbox{
\centering
\includegraphics[width=0.35\textwidth,angle=270]{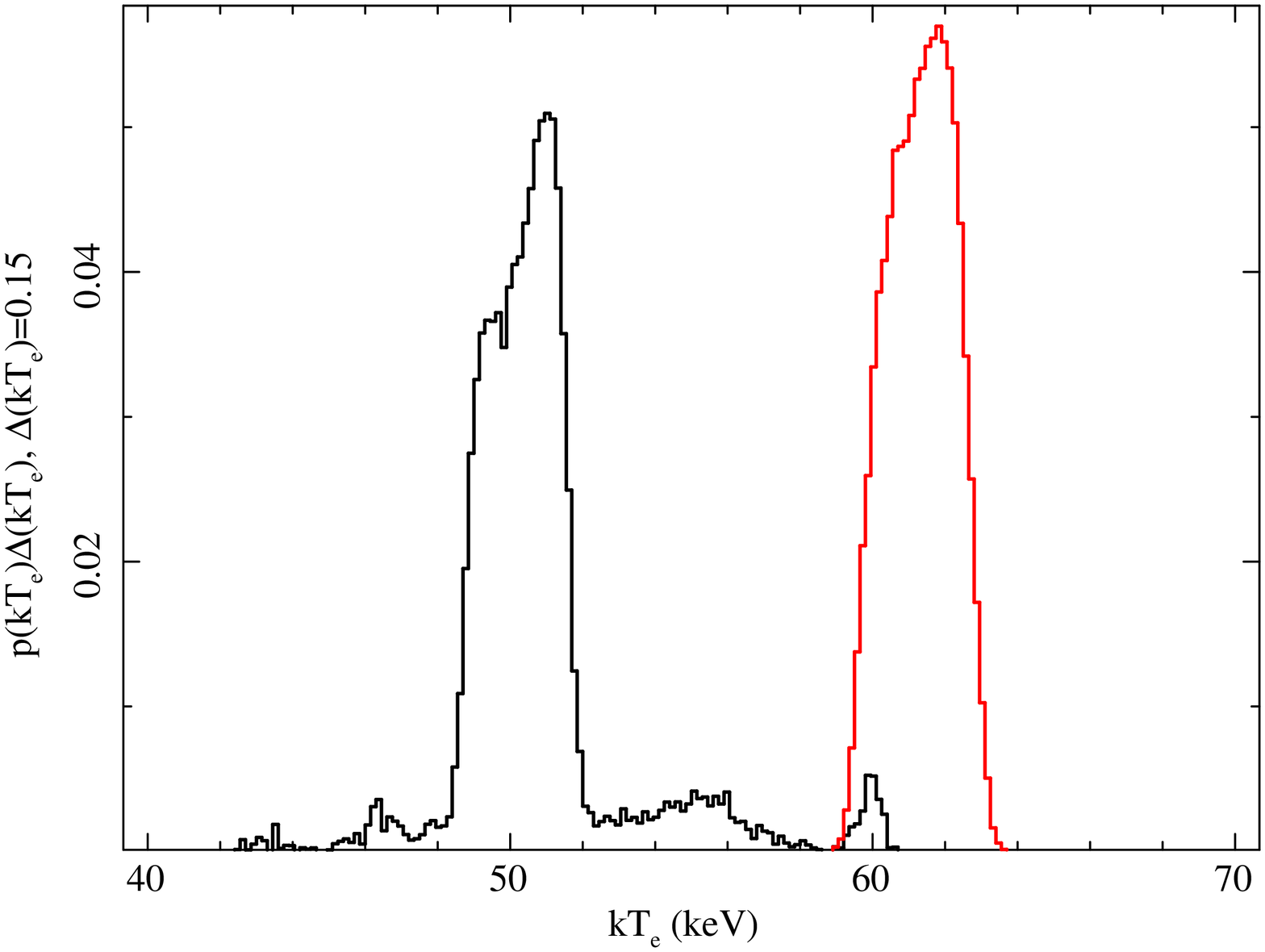}
\includegraphics[width=0.35\textwidth,angle=270]{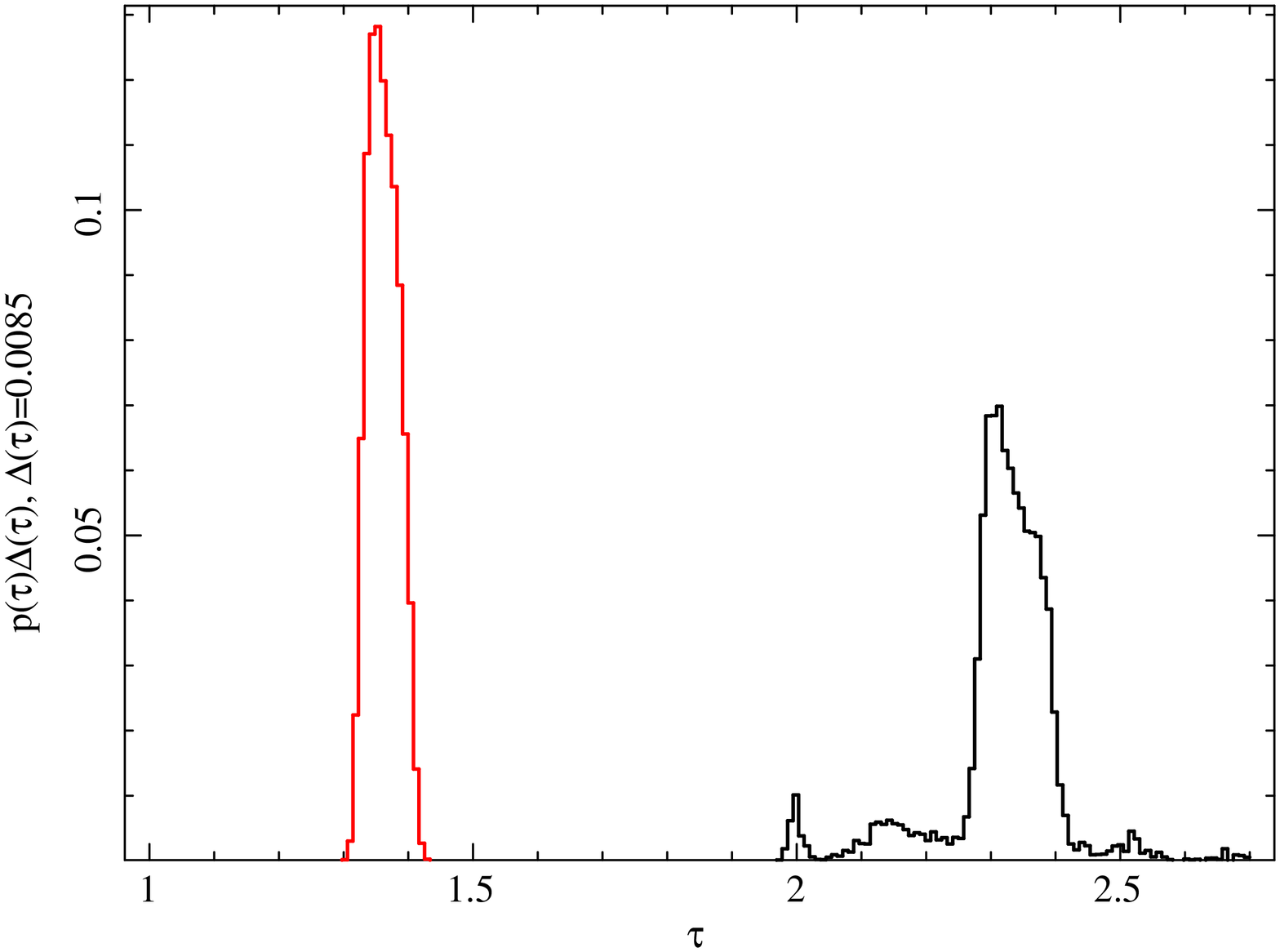}
}
\caption{{\small {\it Left:} Probability density for the {\tt compTT} electron
    temperature ($kT_{\rm e}$) in the spherical (black) and slab (red)
    geometries, as determined from our MCMC analysis.  {\it Right:}
    The same plot, this time for the optical depth ($\tau$) of the
    electrons.  The optical depths for the slab geometry have been
    multiplied by a factor of two (see paper I) to make them more
    directly comparable to the optical depths measured in the
    spherical geometry.}}
\label{fig:hist}
\end{figure}

The goodness-of-fit is largely insensitive to the coronal geometry assumed,
though the temperatures derived from the spherical and slab
geometries are inconsistent at the $>2\sigma$ level ($kT=50^{+6}_{-3}
\keV$ and $kT=61 \pm 1 \keV$, respectively).  The two models also produce different
values for the optical depth of the corona:
$\tau=2.34^{+0.16}_{-0.21}$ and $\tau=0.68 \pm 0.02$,
respectively.  These values differ by $>4\sigma$ (see
Fig.~\ref{fig:hist}) even after the factor-of-two geometrical
difference in calculating the optical depth is accounted for (see
paper I), indicating that a physical change in the properties
of the plasma is necessary when applying a different geometry in order
to achieve a good fit.  This was not the case when Models~2-3 were fit
to the {\it NuSTAR} data alone in paper I.  Unfortunately, the lack of
both short timescale variability and significant relativistic, inner
disk reflection in IC~4329A during our
observing campaign prevents us from determining the
distance of the corona from the accretion disk, and from being able to
constrain how centrally concentrated the coronal emission is.  

Though the spherical
model returns a slightly better statistical fit, we conclude that the
slab model is more physically believable due to the tighter match it
provides between coronal temperature and
power-law cutoff energy (assuming $E_{\rm cut} \sim 2-3kT$; in this case,
$3kT=183 \pm 3 \keV$, which is much closer than Model~2 to being compatible
with the measured power-law cutoff value in Model~1).  We also note
that the temperature and optical depth of the plasma are much more
tightly constrained in the slab geometry.  
Due to an inherent modeling degeneracy between the
optical depth and temperature of the electron plasma in each geometry, however,
there is a small, linearly correlated
range of values for these parameters which demonstrate approximately
equal statistical fit quality, as can be seen in
Figs.~\ref{fig:mo2_contours}a and \ref{fig:mo3_contours}a.  It is thus not
surprising that, as the temperature increases in the slab vs. sphere
case, the optical depth decreases to compensate and produce an
equivalent goodness-of-fit.  Nonetheless, both parameters are
constrained with the best precision and accuracy ever achieved.  

We also note that both sets of values for the coronal temperature and optical
depth measured with the combined, simultaneous {\it Suzaku} and {\it
  NuSTAR} datasets deviate significantly from those obtained when
fitting the {\it NuSTAR} data alone in Models~2-3 of paper I
(Figs.~\ref{fig:mo2_contours}a, \ref{fig:mo3_contours}a and \ref{fig:hist}).  In
particular, the slab
geometry in paper I returned $kT_{\rm e}=37^{+7}_{-6} \keV$
and $\tau=1.25^{+0.20}_{-0.10}$, which is inconsistent with the results
for this geometry using the combined dataset at a $>3\sigma$ level.
We attribute this difference to the larger spectral energy coverage of
the combined dataset, which, as noted previously, allows us to
definitively disentangle the signatures of the continuum, reflection
and absorption in ways that {\it NuSTAR} alone cannot, since its
effective area only extends down to $3 \keV$ and its spectral
resolution in the Fe K band is three times worse than that of {\it
  Suzaku} ($450 \eV$ vs. $150 \eV$).  The amount of reflection in the
system and the curvature induced by the high-energy cutoff of the
continuum are particularly degenerate at energies $>10 \keV$, but
having the high-S/N, high spectral resolution {\it Suzaku} data in the
Fe K band, especially, allows us to
break this degeneracy and to independently constrain $kT_{\rm e}$ and
$K_{\rm refl}$ (see Figs.~\ref{fig:mo2_contours}b and \ref{fig:mo3_contours}b).
Taking these factors into
account, we consider the values for the coronal temperature and
optical depth measured in this work to be the definitive physical
properties of the corona in IC~4329A.  Their deviation from those
determined through the analysis of only $>3 \keV$ data at lower
spectral resolution underscore the importance of obtaining high-S/N
data across a broad X-ray bandpass in order to draw conclusions about
the corona from Comptonization models.

As more constraints on coronal parameters are measured from a sample of AGN, it
will be interesting to compare the coronal properties (e.g., $\Gamma$,
$kT$, $\tau$) with the those of the
black hole and inner accretion flow (e.g., $M_{\rm BH}$, $a$, $\dot{m}$).  It has
long been thought that more actively accreting black holes cool their coronae
more efficiently (\markcite{Skipper2013}{Skipper}, {Mc Hardy}, \&  {Maccarone} 2013 and references therein), but a
sample of AGN with sensitive, broad-band X-ray spectra, as presented
here for IC~4329A, would help to test this conjecture.

\subsection{The Fe K Region}
\label{sec:iron}

The addition of the {\it Suzaku} data enables us to perform a detailed
analysis of the Fe K region in IC~4329A, while the {\it NuSTAR} data
provide an important check on the physical consistency of our models
by simultaneously showing us the Compton reflection continuum $>10
\keV$.  The majority of the reflection component originates in
material at large distances from the black hole (e.g., the putative
torus), and is well fit by a static, neutral {\tt xillver} component.  Though
there is evidence for a broad Fe K$\alpha$ line or a Compton shoulder
due to the residuals remaining after a narrow Fe K$\alpha$ line is
included, the broad line explanation is more likely since the reflection
models we employ already incorporate a Compton shoulder component.
That said, the Compton shoulder explanation cannot be
conclusively ruled out with these data.  

Assuming that the residuals
do correspond to a broad iron line component, modeling this emission feature with a
relativistic line profile (e.g., {\tt diskline, laor} or {\tt
  relline}) yields no improvement in the fit and the model parameters
cannot be constrained.  We have successfully modeled this residual
emission with a Gaussian line at $E \sim 6.4 \keV$, and can place
a limit on its strength relative to the continuum of $EW=24-42
\eV$ (Model~1).
Only $\sim1\%$ of the reflected emission arises from the
broadened Fe K$\alpha$ feature.  While obviously present and
originating from well within the broad
emission line region ($v_{\rm FWHM} \sim 36,000 \kmps$), this feature
likely represents only a weak broad line from the
inner disk.  Indeed, the reflection fraction constrained via the {\tt
  pexrav} model in \S\ref{sec:suzaku} is also low by comparison with other
bright, nearby Seyfert 1 AGN \markcite{Walton2013}({Walton} {et~al.} 2013).
Such a finding is in keeping with the theoretical work of
\markcite{Ballantyne2010}{Ballantyne} (2010), however, who suggested that the majority of Seyferts
may have broad Fe K$\alpha$ lines with $EW \leq 100 \eV$.  Our
inability to constrain any of the parameters
when a relativistic disk line model is applied renders it
useless in constraining the spin of the black hole, however.  Similarly,
attempting to fit this feature with a relativistic smearing kernel
convolved with an ionized disk reflection spectrum also results in no statistical
improvement in fit and no parameter constraints.  The feature is not apparent in
the high-low flux difference spectrum of the AGN, meaning that it is
not significantly variable over the course of the observations.  Even if it does
arise from inner disk reflection this is not surprising, given the lack of short
timescale variability of the continuum.  

A broad Fe
K line has been reported in every observation taken of IC~4329A 
with an X-ray observatory capable of spectrally resolving it
\markcite{Piro1990,Madejski1995,Cappi1996,Perola1999,Done2000,Gondoin2001,McKernan2004,Steenbrugge2005,Markowitz2006,Dadina2007,Molina2009,Molina2013}({Piro}
{et~al.} 1990; {Madejski} {et~al.} 1995; {Cappi} {et~al.} 1996;
{Perola} {et~al.} 1999; {Done} {et~al.} 2000; {Gondoin} {et~al.} 2001;
{McKernan} \& {Yaqoob} 2004; {Steenbrugge} {et~al.} 2005; {Markowitz}
{et~al.} 2006; {Dadina} 2007; {Molina} {et~al.} 2009, 2013). 
Such a line was also noted in the {\it XMM-Newton}
analysis of the source by
\markcite{dlCP2010}{de La Calle P{\'e}rez} {et~al.} (2010), suggesting that this line, though difficult to
characterize definitively, is a persistent
feature of the spectrum over years-long timescales.
Provided that sufficient photon counts have been obtained in the
observation (i.e., $\geq200,000$ from $2-10 \keV$), broad Fe K lines
are detected in $\geq40\%$ of all AGN
\markcite{Guainazzi2006,Nandra2007,dlCP2010}({Guainazzi}, {Bianchi}, \& {Dov{\v  c}iak} 2006; {Nandra} {et~al.} 2007; {de La Calle P{\'e}rez} {et~al.} 2010).  Further, some actively accreting AGN have had broad Fe K
emission lines reported in previous epochs but not currently (e.g.,
NGC~5548, \markcite{Brenneman2012}{Brenneman} {et~al.} 2012).  Taking these points into
consideration, it is perhaps not surprising to find that
IC~4329A does not exhibit strong relativistic reflection signatures during our observation.
Indeed, marginal detections of broad Fe K emission lines such as that
found here may be the norm rather than the exception among even
actively accreting AGN \markcite{Ballantyne2010}({Ballantyne} 2010).  Within this framework,
it is intriguing to
note that the source is accreting at $L_{\rm bol}/L_{\rm Edd} \sim
0.46$ for a black hole with an estimated mass of $M_{\rm BH}=1.20 \times 10^8 \Msun$
\markcite{dlCP2010}({de La Calle P{\'e}rez} {et~al.} 2010).  Given that the Keplerian velocity of the broadened Fe K$\alpha$
feature places its origin at $r \sim 70\,r_{\rm g}$ from the black
hole, this suggests that the optically thick disk may not extend down to the
ISCO.  The disk may be truncated within this radius, or perhaps it is too
highly ionized to significantly contribute to the reflection
spectrum.  Indeed, highly ionized disks are expected in relatively
high accretion rate sources \markcite{Ballantyne2011}({Ballantyne}, {McDuffie}, \&  {Rusin} 2011) such as IC~4329A.
The power-law photon index of the source is also
considerably harder ($\Gamma \sim 1.73$) than is typical for an
actively accreting source with an inner disk extending down to its
ISCO, and marks a departure of $>6\sigma$ from the AGN relation measured by
\markcite{Brightman2013}{Brightman} {et~al.} (2013).  According to these authors, for an Eddington
ratio of $L_{\rm bol}/L_{\rm Edd}=0.46$, one should measure
$\Gamma=2.16 \pm 0.07$, in contrast to the $\Gamma=1.73 \pm 0.01$
measured here for IC~4329A (however, the intrinsic scatter in this
relation must be considered, as must the uncertainty in measuring the
Eddington ratio in a given source).

The relative weakness of the reflection features compared to
similar AGN coupled with the hard power-law index of the source,
particularly, lends credence to the hypothesis put forward in paper I:
that we are witnessing an outflowing corona with $v_{\rm out} \sim
0.2c$, following the work of
\markcite{Beloborodov1999}{Beloborodov} (1999) and \markcite{Malzac2001}{Malzac}, {Beloborodov}, \&  {Poutanen} (2001).  Although an ionized
inner disk would certainly inhibit strong reflection features from
this region, as per \markcite{Ballantyne2010}{Ballantyne} (2010), it is worth noting that
the outflowing corona scenario would suppress them as well: if the
main locus of coronal emission is situated at a height of
$\geq50\,r_{\rm g}$ then we become insensitive to reflection from the
disk within $50\,r_{\rm g}$.  Also, if the corona is relativistically
outflowing then aberration decreases the illumination of the inner
disk, again making us less sensitive to any reflection from this region.
Under any of the above conditions we would not expect to be able to
constrain the spin of the black hole in IC~4329A.
A deep multi-wavelength campaign involving UV spectra, particularly, in
addition to the outstanding data now available in X-rays with {\it
  NuSTAR} and {\it Suzaku, XMM-Newton} or {\it Chandra} would be
necessary in order to properly evaluate the characteristics and
structure of the inner accretion disk, and to place our results on the
energetics of the system in their proper context.

\acknowledgements{This work was supported under NASA Contract No. NNG08FD60C, and
made use of data from the {\it NuSTAR} mission, a project led by
the California Institute of Technology, managed by the Jet Propulsion
Laboratory, and funded by the National Aeronautics and Space
Administration. We thank the {\it NuSTAR} Operations, Software and
Calibration teams for support with the execution and analysis of
these observations.  This research has made use of the {\it NuSTAR}
Data Analysis Software (NuSTARDAS) jointly developed by the ASI
Science Data Center (ASDC, Italy) and the California Institute of
Technology (USA).  LB thanks Koji Mukai and the {\it Suzaku} GOF at NASA/GSFC for
  all their assistance in obtaining and analyzing that data, and also gratefully
  acknowledges funding from NASA grant NNX13AE90G.  G.~Matt and AM
  acknowledge financial support from Italian
Space Agency under contract ASI/INAF I/037/12/0 - 011/13.}

\bigskip


\bibliographystyle{apj}

\end{document}